\documentclass[5p,times,twocolumn,a4paper,showpacs,showkeys,amsmath,amssymb,superscriptaddress,nofootinbib]{elsarticle}

\usepackage{hyperref}

\usepackage{lineno}
\usepackage{comment}
\usepackage{upgreek}
\biboptions{sort&compress}
\usepackage{siunitx}
\usepackage{mathtools}
\usepackage{physics}
\usepackage{adjustbox}
\AtBeginDocument{\RenewCommandCopy\qty\SI}
\usepackage{tikz}
\usetikzlibrary{arrows.meta, calc}

\usepackage{float}
\usepackage[small,bf]{caption}
\usepackage[subrefformat=parens]{subcaption}
\usepackage[whole]{bxcjkjatype} % Just used to discuss the paper's structure in Japanese 
\usepackage{xcolor} % For text color
\usepackage{colortbl}
\usepackage{graphicx}
\graphicspath{{figure}}

\modulolinenumbers[5]

\newcommand{\figref}[1]{Fig.~\ref{#1}}
\newcommand{\tabref}[1]{Table~\ref{#1}}
\renewcommand{\eqref}[1]{Eq.~\ref{#1}}
\newcommand{\secref}[1]{Sec.~\ref{#1}}
\newcommand{\citeref}[1]{Ref.~\cite{#1}}
\DeclareRobustCommand{\upe}{\ifmmode{\mathrm{e}}\else{e}\fi}
\DeclareRobustCommand{\meg}{\ifmmode{\mu^+ \to e^+ \gamma}\else{\(\mu^+ \to e^+ \gamma~\)}\fi}
\newcommand{\eff}[1]{\varepsilon_{\mathrm{#1}}} %% efficiency of #1
\newcommand{\tleft}{t_\mathrm{left}}
\newcommand{\tright}{t_\mathrm{right}}

\newcommand{\prototype}{converter prototype}
\newcommand{\lysodimension}[3]{\SI{#1}{mm \times} \SI{#2}{mm \times} \SI{#3}{mm}~(thickness)}

\begin{document}


% \title{Light yield and time resolution of LYSO crystal as an active converter material for 52.8 MeV photon detection in future $\meg$ search} % Original
\title{Performance of an LYSO-Based Active Converter for a Conversion Spectrometer aiming for \SI{52.8}{MeV} photon detection in Future $\meg$ Search Experiments} % Wataru's suggestion
% \title{Performance of an LYSO-Based Active Converter for Detecting \SI{52.8}{MeV} Photons in a Pair Spectrometer for Future $\meg$ Experiments} 

\cortext[mycorrespondingauthor]{Corresponding author}
\author[icepp]{Sei Ban}
\author[icepp]{Lukas Gerritzen}
\author[u-tokyo]{Fumihito Ikeda}
\author[icepp]{Toshiyuki Iwamoto}
\author[icepp]{Wataru Ootani}
\author[icepp]{Atsushi Oya} 
\ead{atsushi@icepp.s.u-tokyo.ac.jp}
\author[u-tokyo]{Rei Sakakibara\corref{mycorrespondingauthor}}
\ead{rei@icepp.s.u-tokyo.ac.jp}
\author[u-tokyo]{Rintaro Yokota}

\address[icepp]{International Center for Elementary Particle Physics, The University of Tokyo, Bunkyo-ku, Tokyo 113-0033, Japan}
\address[u-tokyo]{Department of Physics, The University of Tokyo, Bunkyo-ku, Tokyo 113-0033, Japan}

\begin{frontmatter}

\begin{abstract}
\noindent To facilitate future \meg search experiments with a branching‐ratio sensitivity of \num{e-15}, 
we are developing a conversion spectrometer that incorporates an active LYSO converter.
The converter generates $e^+e^-$ pairs from incident photons while simultaneously measuring their energy deposition and timing, 
thereby enabling precise reconstruction of \SI{52.8}{MeV} photons.
The design goals include a time resolution of \SI{30}{ps} and an energy resolution of \SI{200}{keV} for the detection of \SI{52.8}{MeV} photons.
Based on simulation studies, we optimized the converter thickness and segment dimensions, followed by the  fabrication of prototype LYSO segments.
The single-MIP detection performance of these prototypes was evaluated using an electron beam at the KEK PF-AR test beamline.
The prototypes demonstrated excellent performance, achieving a time resolution of \SI{25}{ps} and a light yield of $10^4$ photoelectrons, both of which significantly exceed the design requirements.
\end{abstract}

\begin{keyword}
%\texttt{elsarticle.cls}\sep \LaTeX\sep Elsevier \sep template
%\MSC[2010] 00-01\sep  99-00
$\meg$ experiment, conversion spectrometer, LYSO
\end{keyword}

\end{frontmatter}

% \linenumbers
\section{Introduction}\label{sec:Introduction}
\noindent A highly sensitive search for the charged lepton flavor-violating decay \meg provides a powerful probe for physics beyond the Standard Model (SM).
In the SM, this decay is forbidden by lepton flavor conservation.
Though it can proceed via neutrino masses and mixing, its branching ratio is negligibly small ($\sim \num{e-54}$) \cite{Petcov:1976ff, KUNONuOscMEG}. 
In contrast, various new physics models predict branching ratios within an experimentally accessible range of \SIrange{e-14}{e-12}{} \cite{Barbieri:1994pv, Hisano:1995cp, Hisano:1996qq, Hisano:1998cx, Hisano:1998fj, Antusch:2006vw, Calibbi:2006nq, Moroi:2013vya, Hirao:2021pmh}.
\par
\indent Experimentally, \meg is identified by its distinct two-body kinematics: 
the positron and photon each carry an energy of \SI{52.8}{MeV} and are emitted simultaneously and back-to-back.
The background predominantly arises from accidental overlaps of a positron and a photon originating from separate muon decays. 
The number of such background events scales as: 
\begin{equation}
   \label{eq:NAcc}
      N_\mathrm{acc}\propto R_\mu^2\ \Delta t_{e\gamma}\ (\Delta E_\gamma)^2\ \Delta E_e\ (\Delta \Theta_{e\gamma})^2,
\end{equation}
where $R_\mu$ is the muon beam rate, and $\Delta$ denotes the resolutions for the photon and positron energies
($E_\gamma$ and $E_e$), their relative timing ($t_{e\gamma}$), and the opening angle between them ($\Theta_{e\gamma}$).
To suppress this background, a continuous muon beam and high-resolution detectors are indispensable.
The MEG~II experiment is currently underway at the Paul Scherrer Institute (PSI) $\pi E5$ beam line \cite{PiE5}, 
which delivers the world's most intense continuous muon beam with rates up to \SI{e8}{\mu^+\per\second}.
Following the latest experimental limit on the branching ratio of \num{1.5e-13} (\SI{90}{\percent} C.L.) \cite{MEGII2022Result}, 
MEG~II aims to reach a sensitivity of \num{6e-14} by the conclusion of physics data-taking in 2026 \cite{MEGIIDetectorPaper}.
Looking further ahead, PSI plans to construct a new muon beam line, the High-Intensity Muon Beam (HIMB), in \SIrange{2027}{2028}{} 
\cite{IMPACT_HIMB, SCIENCE_HIMB}.
With muon rates reaching \SI{e10}{\mu^+\per\second}, HIMB will open the door for next-generation experiments targeting sensitivities 
at the $\mathcal{O}(10^{-15})$ level \cite{LOI2026}.
Realizing such sensitivities requires detectors with unprecedented performance.
In this paper, we describe the development of a novel photon detector utilizing a conversion spectrometer with 
an active LYSO converter, designed to achieve energy and timing resolutions of $\Delta E_\gamma < \SI{200}{keV}$ and 
$\Delta t_\gamma < \SI{30}{ps}$ at \SI{52.8}{MeV}.
\par
\indent In the earlier MEGA experiment\cite{MEGAExperiment}, photon detection was performed using a conversion spectrometer where photons were converted 
into $e^+e^-$ pairs within thin, passive converter layers.
In such a passive converter, the photon energy is reconstructed solely from the measured $e^+e^-$ momenta.
However, the precision is inherently limited by energy losses within the converter material—primarily via ionization 
and bremsstrahlung—which cannot be fully recovered on an event-by-event basis.
To overcome this fundamental limitation, we propose an active converter material that enables the direct measurement 
of ionization energy loss within the converter itself.
By integrating this measurement with the reconstructed track momenta, the photon energy resolution can be significantly 
enhanced beyond the capabilities of passive systems.
\par
\indent The remainder of this paper is organized as follows. 
\secref{sec:DesignConcept} outlines the design concept. 
\secref{sec:Simulation} details simulation studies of the photon conversion and $e^+e^-$ tracking, 
defining the light yield requirements and design optimization.
\secref{sec:BeamTest} describes the evaluation of the LYSO active converter prototypes using an electron beam.
The results for timing performance and light yield are reported in \secref{sec:TimingPerformance} and \secref{sec:LightYield}, 
respectively.
Finally, these results are discussed in \secref{sec:Discussion}, followed by conclusions in \secref{sec:Conclusion}.

%%%%%%%%%%%%%%%%%%%%%%%%%%%%%%%%%%%%
%%
%% Design concept section
%%
%%%%%%%%%%%%%%%%%%%%%%%%%%%%%%%%%%%%
\section{Concept of a conversion spectrometer for future \meg searches}\label{sec:DesignConcept}
\begin{figure}[tbp]
   \centering
   \includegraphics[width = 1.\linewidth]{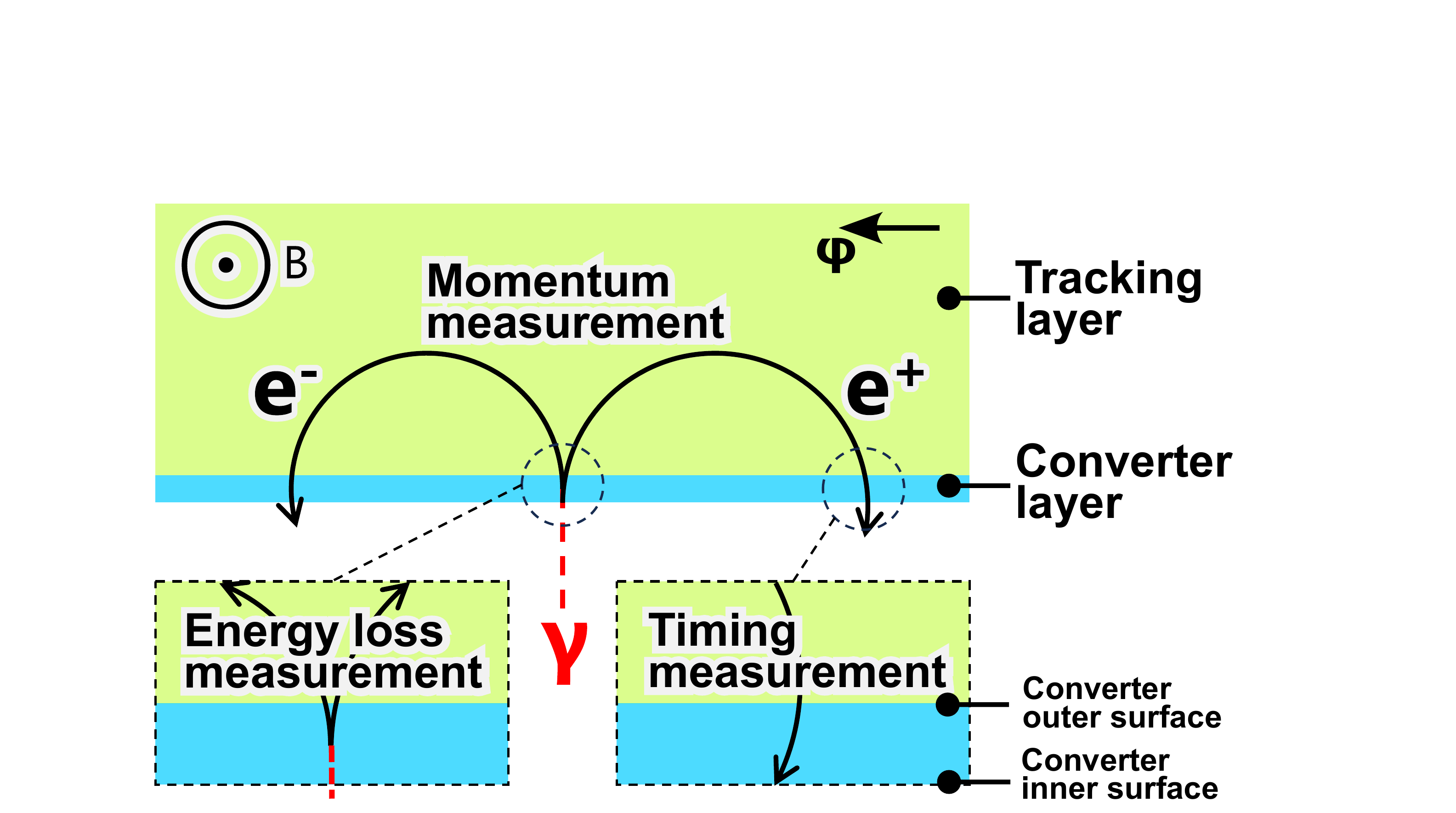}
   \caption{
      Principle of energy and timing measurement using a conversion spectrometer with an active converter.
      The symbol $\phi$ denotes the azimuthal angle in the detector coordinate system defined in \figref{fig:fullsketch}.
      The arrow $B$ indicates the direction of the magnetic field.
   }
   \label{fig:pairspectrometer}
\end{figure}
\begin{figure}[tbp]
    \centering
    \includegraphics[width = 1.\linewidth]{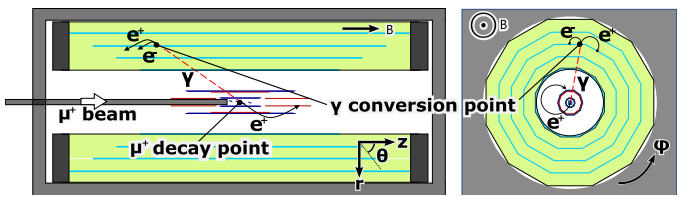}
    \caption{
      A schematic overview of a possible future \meg experiment detector layout.
      A magnetic field is applied in the $z$-axis, defined in the figure.
   }
    \label{fig:fullsketch}
\end{figure}

\noindent As introduced in \secref{sec:Introduction}, a conversion spectrometer employing an active converter is a candidate technology for achieving the high resolution required for \SI{52.8}{MeV} photon measurements.
An incident photon is converted into an $e^+e^-$ pair within a thin conversion layer, and both tracks are precisely measured by a pair-tracking layer 
coupled to the converter (\figref{fig:pairspectrometer}).
Here, we define a cylindrical coordinate system for the detector, where the $z$-axis is aligned with the muon beam direction (as shown in \figref{fig:fullsketch}).
A magnetic field $B$ is applied along the $z$-axis, and $\phi$ represents the azimuthal angle.
\par
\indent We utilize a crystal scintillator as the converter material, as it provides a high conversion probability while enabling the measurement of energy loss by 
$e^+e^-$ pairs traversing the converter.
The total energy of an incident photon is expressed as:
\begin{equation} \label{eq:EGammaReco}
   E_\gamma = E_{e^+} + E_{e^-} + E_\mathrm{dep} + E_\mathrm{rad},
\end{equation}
where $E_{e^\pm}$ denotes the energy (reconstructed from momenta) measured by the tracking layer, 
and $E_\mathrm{dep}$ is the energy deposition measured in the active converter.
The term $E_\mathrm{rad}$ represents the energy carried away by radiative processes, such as bremsstrahlung photons from the $e^+e^-$ pair, 
which escape detection by the converter.
Although $E_\mathrm{rad}$ cannot be recovered on an event-by-event basis, the active measurement of $E_\mathrm{dep}$ minimizes the unmeasured energy fraction, 
which is essential for achieving superior resolution.
To prevent energy deposits from pileup photons and re-entering tracks (the latter illustrated in \figref{fig:Boomerang} and discussed in \secref{sec:CrystalSegmentation}) 
from biasing the $E_\mathrm{dep}$ measurement, the active converter is segmented.
While large crystals can lead to detection inefficiencies, excessively small segments increase the number of readout channels; 
therefore, the segmentation scheme must be optimized to balance performance and practical feasibility.
The timing of the photon conversion is reconstructed from the hit times of the $e^+e^-$ tracks in the converter.
The converter measures the timing of the $e^+e^-$ tracks at their re-entry points after completing a half-turn ($t_\mathrm{hit}^{e^\pm}$). 
The photon conversion time is then reconstructed by applying time-of-flight corrections ($t_\mathrm{TOF}^{e^\pm}$):
\begin{equation}
   t_\gamma = \frac{1}{2}\left(t_\mathrm{hit}^{e^-}+t_\mathrm{hit}^{e^+}-t_\mathrm{TOF}^{e^-}-t_\mathrm{TOF}^{e^+}\right).
\end{equation}
\par
\indent The target resolutions are \SI{200}{keV} for $E_\gamma$ and \SI{30}{ps} for $t_\gamma$ at \SI{52.8}{MeV}.
To maintain the $E_\mathrm{dep}$ resolution below \SI{200}{keV}, the active converter must provide sufficient light yield, 
as quantified in \secref{sec:Simulation}.
Furthermore, achieving an overall $t_\gamma$ resolution of \SI{30}{ps} requires a resolution of better than \SI{40}{ps} for each $t_\mathrm{hit}^{e^\pm}$, 
assuming a $\sqrt{2}$ scaling from the two independent measurements.

%%%%%%%%%%%%%%%%%%%%%%%%%%%%%%%%%%%%
%% Efficiency (stacking approach)
%%%%%%%%%%%%%%%%%%%%%%%%%%%%%%%%%%%%
\indent Although an excellent resolution can be expected, this design suffers from a low conversion efficiency in the thin converter layers, amounting to only \SI{2.2}{\percent} per single layer, as discussed in \secref{sec:Simulation}.
To mitigate this limitation, a stacking of multiple converter layers is employed, 
and the detector geometry is designed to provide a large geometrical acceptance, as illustrated in \figref{fig:fullsketch}.
As a result, a high acceptance of \SIrange{80}{90}{\percent} can be achieved, which far exceeds that of the MEG~II experiment (\SI{11}{\percent}).
This extensive acceptance, particularly in the polar angle $\theta$, enables the measurement of the photon emission angular distribution.
When combined with a polarized muon beam, this capability allows discrimination among physics models beyond the Standard Model in the event of a discovery \cite{KunoOkadaReview}.
With its excellent resolution and a muon beam rate of up to \SI{e10}{\mu^+\per\second}, the proposed design is expected to reach a branching-ratio sensitivity of $\mathcal{O}(\num{e-15})$.

%%%%%%%%%%%%%%%%%%%%%%%%%%%%%%%%%%%%
%%
%% Simulation section
%%
%%%%%%%%%%%%%%%%%%%%%%%%%%%%%%%%%%%%
\section{Simulation study of active converter design}\label{sec:Simulation}
\noindent The active converter design was optimized through Geant4-based simulations, primarily to maximize the signal efficiency.
In \secref{sec:EfficiencySimulation}, we define the benchmark signal efficiency and present its value for the baseline design.
We also describe three detector effects considered in the design optimization: the conversion probability, the energy loss of $e^+e^-$ tracks inside the converter, and the topology of the $e^+e^-$ tracks.
In \secref{sec:MaterialAndThickness}, we discuss the optimization of the converter material and its thickness, taking into account the first two effects.
In \secref{sec:CrystalSegmentation}, we discuss the converter segmentation, which is relevant to both pileup effects and inefficiencies arising from the third detector effect.
Based on the optimized design, \secref{sec:ConverterLYrequirement} quantifies the required light yield of the converter.
Throughout the simulation, a uniform magnetic field of \SI{2}{T} is assumed in the pair-tracker, although this value has not yet been finalized.

%%%%%%%%%%%%%%%%%%%%%%%%%%%%%%%%%%%%%%%%
%% signal efficiency characterization %%
%%%%%%%%%%%%%%%%%%%%%%%%%%%%%%%%%%%%%%%%
\subsection{Characterization of conversion spectrometer performance}\label{sec:EfficiencySimulation}
\begin{figure}[tbp]
   \centering
   \includegraphics[width = 0.7\linewidth] {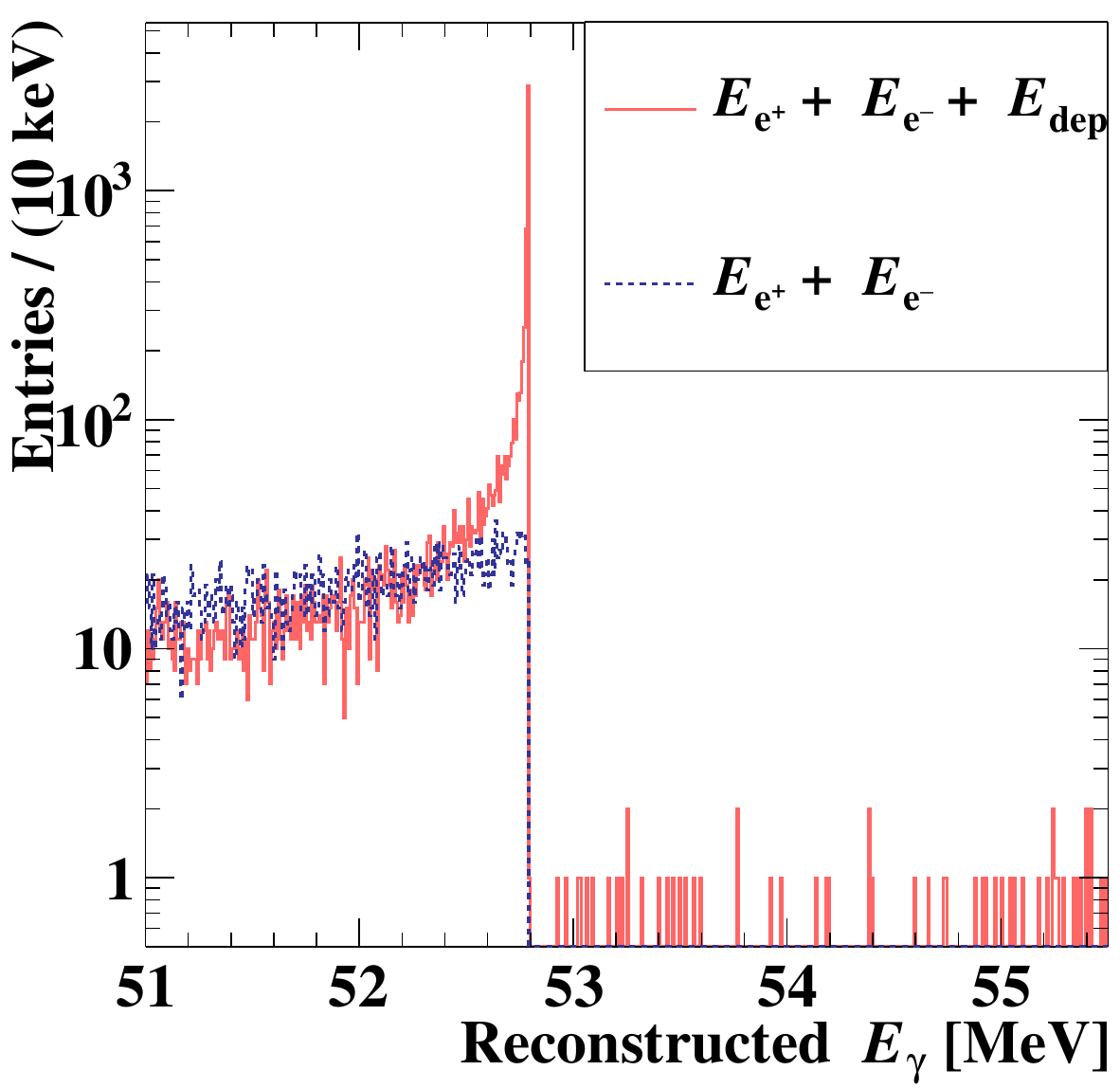}
   \caption{
      Reconstructed spectrum of the simulated signal photon energies, assuming a \SI{3}{mm}-thick LYSO converter.
      The blue dotted line shows the momentum sum of the $e^+e^-$ pair in the pair-tracker.
      The red line represents the reconstructed energy obtained by adding the energy deposited in the converter to the momentum sum.
   }
   \label{fig:EGammaRec}
\end{figure}

\noindent To characterize the performance of the active converter design, we evaluated the energy spectrum of signal photons reconstructed from the converter and pair-tracker information.
The reconstructed photon energy $E_\gamma$ is defined as the sum of the $e^+e^-$ momenta measured in the pair-tracker and the energy deposited in the LYSO converter ($E_\mathrm{dep}$), where $E_\mathrm{dep}$ is calculated by integrating the energy deposits within a typical scintillation-light integration time.
In this performance study, while $e^\pm$ momenta and $E_\mathrm{dep}$ are based on Monte Carlo truth information to decouple detector resolution effects, $E_\mathrm{dep}$ inherently includes additional deposits from the same $e^\pm$ tracks if they traverse the crystal multiple times along their helical trajectories.
\figref{fig:EGammaRec} shows the resulting energy spectra with and without the $E_\mathrm{dep}$ measurement.
The inclusion of $E_\mathrm{dep}$ significantly improves the energy reconstruction, shifting the distribution to a sharp peak at \SI{52.8}{MeV}.
This allows for effective discrimination against background photons, although a low-energy tail persists due to the unrecoverable energy component $E_\mathrm{rad}$ described in \secref{sec:DesignConcept}.
\par
\indent Based on this spectrum, the benchmark signal efficiency, $\eff{sig}$, was evaluated by counting the number of photons within the signal energy window of $\SI{52.7}{MeV} < E_{\gamma} < \SI{52.9}{MeV}$, a range determined by the target energy resolution of \SI{200}{keV}.
This efficiency can be factorized into four terms:
\begin{equation}\label{eq:efficiency}
   \eff{sig} = \eff{geom} ~ \eff{conv} ~ \eff{topo} ~ \eff{analysis}.
\end{equation}
\noindent The first term, $\eff{geom}$, represents the geometrical acceptance.
The second term, $\eff{conv}$, represents the efficiency determined by the underlying physical processes within the active converter.
This term accounts for two contributions:
the probability that a photon converts inside the converter and that both $e^+e^-$ properly exit the converter,
and the loss of converted events in which bremsstrahlung carries away energy, causing the reconstructed photon energy to populate the low‑energy tail and fall outside the signal window.
The third term, $\eff{topo}$, represents the efficiency associated with the event topology.
The $e^+e^-$ tracks may traverse the initial crystal multiple times along their helical trajectories in the magnetic field.
In such cases, the active converter records additional energy deposits during subsequent crossings, leading to an overestimation of $E_\mathrm{dep}$.
If these events can be identified from the reconstructed tracks in the pair-tracker, they are excluded from the analysis. 
Otherwise, they contribute to the high-energy tail in \figref{fig:EGammaRec} and increase the background contamination around 
\SI{52.8}{MeV} (see \figref{fig:RMDspectrum}), which reduces signal efficiency.
The fourth term, $\eff{analysis}$, represents the analysis-related efficiencies not included in the previous terms.
While a realistic evaluation requires a detailed design of the spectrometer components and beam conditions, we applied the following benchmark selection criteria.
Vertex matching between $e^+e^-$ tracks (within \SI{2}{mm}) was required at the Monte Carlo truth level, which reduced $\eff{analysis}$ by \SI{1}{\percent}.
Additionally, $e^+e^-$ tracks with a transverse momentum below \SI{5}{MeV} were rejected, as they are expected to be poorly reconstructed.
Accounting for these factors, $\eff{analysis}$ is estimated to be \SI{85}{\percent}.
\par
\indent The efficiency components for the baseline converter design are summarized in \tabref{tab:EfficiencyBreakdown}.
The design employs a LYSO crystal with dimensions of \SI{50}{mm}, \SI{5}{mm}, and \SI{3}{mm} in the $z$, azimuthal, and radial directions, respectively; further details are provided in \secref{sec:MaterialAndThickness} and \secref{sec:CrystalSegmentation}.
These values are calculated sequentially; for instance, $\eff{topo}$ is defined as the fraction of events with acceptable topology among those that satisfy the $\eff{conv}$ criteria.

\begin{table}[tbp]
    \centering
    \begin{tabular}{c||cccc}\hline
       $\eff{sig}$        & $\eff{geom}$        & $\eff{conv}$        &  $\eff{topo}$       & $\eff{analysis}$   \\\hline
       \SI{1.9}{\percent} & \SI{85}{\percent}  & \SI{2.7}{\percent}  & \SI{95.6}{\percent} & \SI{85}{\percent}  \\\hline
    \end{tabular}
    \caption{
   Breakdown of the signal efficiency into its constituent factors for the baseline converter design.
}
    \label{tab:EfficiencyBreakdown}
\end{table}

%%%%%%%%%%%%%%%%%%%%%%%%%%%%
%% Material and thickness %%
%%%%%%%%%%%%%%%%%%%%%%%%%%%%
\subsection{Active converter material and thickness}\label{sec:MaterialAndThickness}
\begin{figure}[tbp]
   \centering
   \includegraphics[width = 0.7\linewidth]{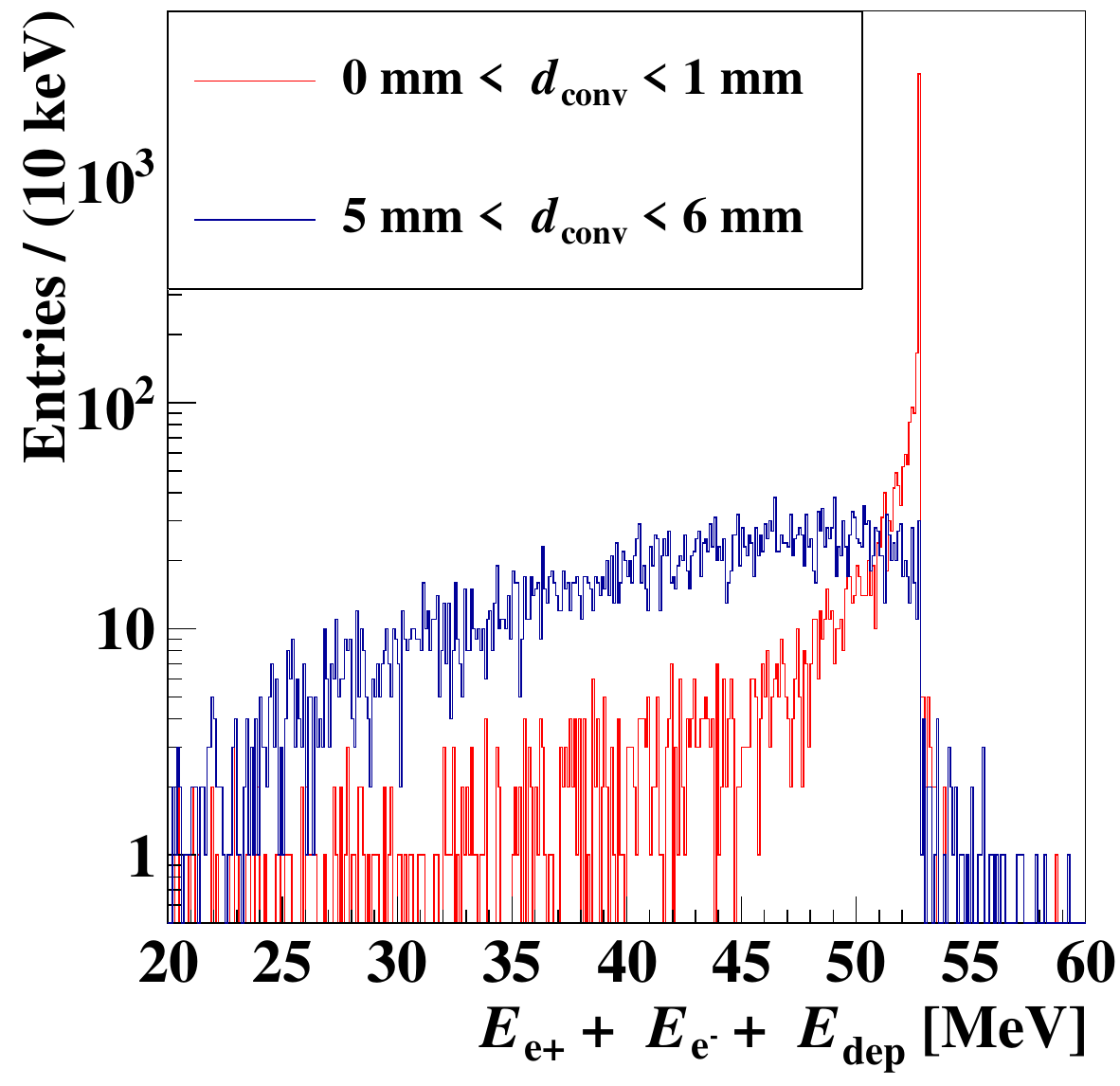}
   \caption{
      Reconstructed $E_\gamma$ spectra for signal photons converted in LYSO at different depths $x$, 
      defined as the distance from the conversion point to the outer surface. 
      The red line corresponds to events with $\SI{0}{mm} < x < \SI{1}{mm}$, while the blue line corresponds to events with $\SI{5}{mm} < x < \SI{6}{mm}$.
   }
   \label{fig:EnergySpectrum_ConversionDepth}
\end{figure}

\noindent The thickness and material of the converter directly affect $\varepsilon_\mathrm{conv}$, 
which can be decomposed into two effects as follows:
\begin{equation}\label{eq:EfficiencyPhysicalProcess}
   \eff{conv} = \int_0^{d_\mathrm{thick}} p_\mathrm{conv}(x;d_\mathrm{thick})~p_\mathrm{peak}(x)~\text{d}x,
\end{equation}
where $d_\mathrm{thick}$ is the thickness of the converter, and $x$ denotes the conversion depth, with $x=0$ corresponding to the outer surface 
(see \figref{fig:pairspectrometer} for the definition).
The term $p_\mathrm{conv}(x;d_\mathrm{thick})$ represents the probability density for a photon to convert at depth $x$ within the converter, 
following:
\begin{equation}\label{eq:ConversionProbability}
   p_\mathrm{conv}(x;d_\mathrm{thick}) \propto \exp\left(\frac{x-d_\mathrm{thick}}{\ell_\mathrm{mean}}\right),
\end{equation}
where $\ell_\mathrm{mean}$ is the mean free path of \SI{52.8}{MeV} photons in the converter.
The second term in \eqref{eq:EfficiencyPhysicalProcess}, $p_\mathrm{peak}(x)$, is the probability that a photon converted at depth $x$ is observed within the 
\SI{52.8}{MeV} peak; this probability is independent of $d_\mathrm{thick}$.
If the dependence of $p_\mathrm{peak}(x)$ is neglected, the conversion probability simply increases with the converter thickness, scaling as $1-\exp(-d_\mathrm{thick}/\ell_\mathrm{mean})$.
However, as discussed in \secref{sec:EfficiencySimulation}, the efficiency is also limited by energy leakage from the converter, 
which predominantly originates from bremsstrahlung photons that escape the converter without detection.
Such leakage increases as the $e^+e^-$ pairs travel a longer path length inside the material.
Furthermore, the energy required for the pair to exit the converter and be successfully measured by the pair-tracker increases with the conversion depth $x$.
Consequently, $p_\mathrm{peak}(x)$ is a decreasing function of $x$.
This effect is illustrated in \figref{fig:EnergySpectrum_ConversionDepth}, which compares the reconstructed $E_\gamma$ spectra for conversions occurring near the outer 
and inner surfaces of the converter.
The spectrum corresponding to deeper conversions exhibits a smaller \SI{52.8}{MeV} peak, implying a reduction in $p_\mathrm{peak}(x)$.
\par
\begin{figure}[tbp]
   \centering
   \includegraphics[width = 0.7\linewidth]{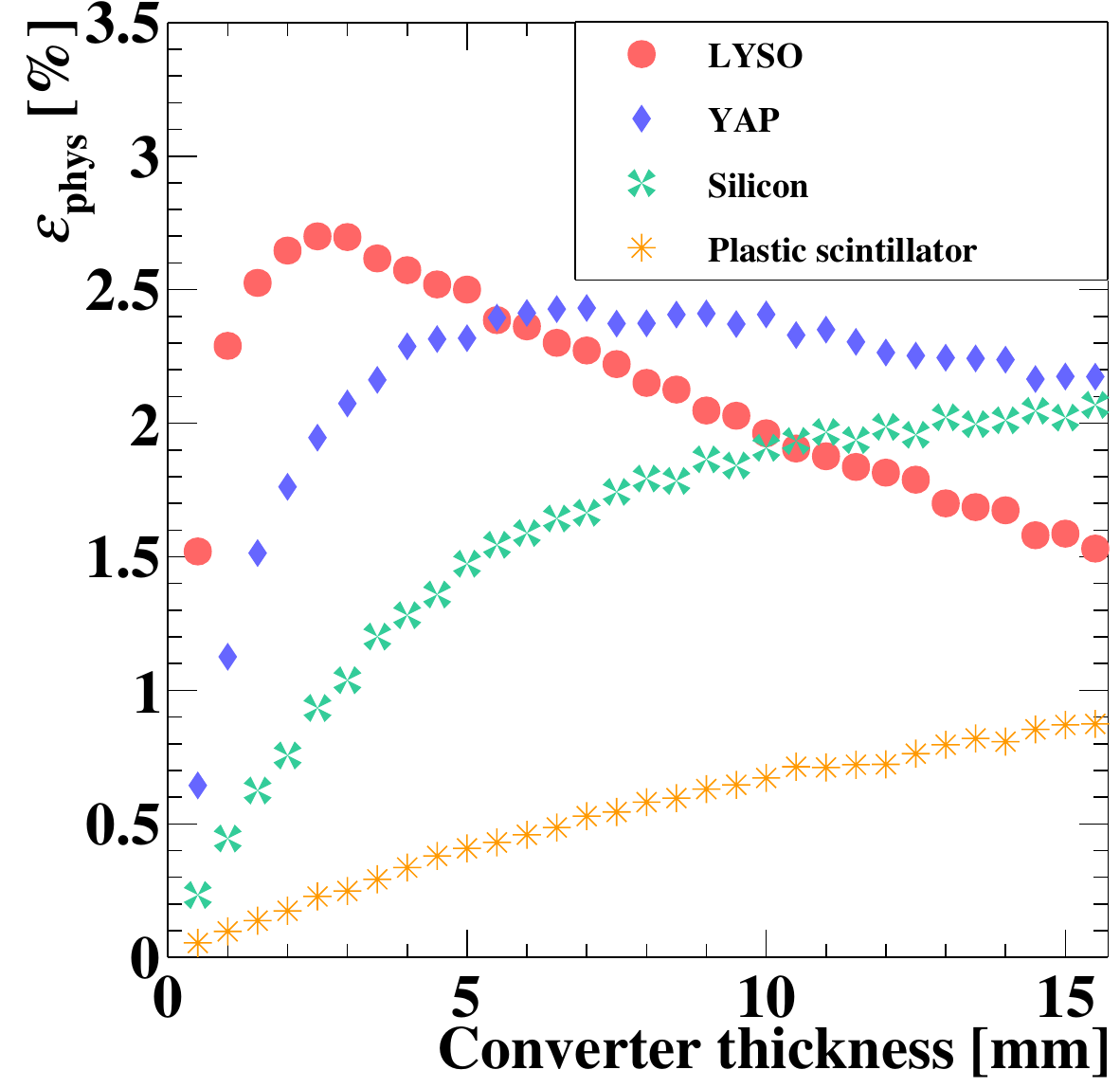}
   \caption{Simulated $\varepsilon_\mathrm{conv}$ value as a function of converter thickness for various materials.}
   \label{fig:converter_material_thickness}
\end{figure}

\indent We performed simulations to compare $\eff{conv}$ for various converter materials and thicknesses, as shown in \figref{fig:converter_material_thickness}.
The material comparison indicates that the signal efficiency is primarily governed by the material density, 
as thin converters made of dense materials provide a sufficient pair-production probability while limiting the total energy loss.
Regarding the thickness dependence, $\eff{conv}$ initially increases with $d_\mathrm{thick}$, then saturates, and eventually begins to decrease.
The initial rise is driven by the increasing conversion probability in the region where $p_\mathrm{peak}(x)$ retains significant values.
The saturation point corresponds to the depth where $p_\mathrm{peak}(x)$ approaches zero, beyond which further increases in thickness do not contribute to $\eff{conv}$.
When the converter is even thicker, $\eff{conv}$ decreases exponentially due to the scaling behavior of $p_\mathrm{conv}(x;d_\mathrm{thick})$ 
described in \eqref{eq:ConversionProbability}.
\par
\indent Among the materials considered, LYSO exhibited the highest $\eff{conv}$, 
reaching \SI{2.7}{\percent} at the efficiency plateau within the thickness range of \SIrange{2.5}{3}{mm}.
Moreover, it possesses excellent scintillation properties, including a high light yield and fast response, 
making it the most suitable choice for an active converter.
The baseline thickness was selected as \SI{3}{mm}, the thicker end of the efficiency plateau, 
where better time resolution is expected.
\par
\indent The photon incident angle on the converter must also be taken into account.
In the above optimization, a thickness of \SI{3}{mm} was found to be optimal under the assumption that photons are incident perpendicular to the converter surface 
($\theta_\gamma = \ang{90}$), where the thickness $d_\mathrm{thick}$ corresponds to the path length along the photon trajectory.
However, in regions with large $|z|$, the typical photon incident angle deviates from \ang{90}, 
resulting in an effective path length of $d_\mathrm{thick}/\sin\theta_\gamma$.
To maintain a consistent path length within the converter independent of $\theta_\gamma$, the converter thickness must be adjusted accordingly.

%%%%%%%%%%%%%%%%%%%%%%%%%%
%% Crystal segmentation %%
%%%%%%%%%%%%%%%%%%%%%%%%%%
\subsection{Active converter segmentation}\label{sec:CrystalSegmentation}
\begin{figure}[tbp]
   \centering
   \begin{minipage}[t]{0.48\linewidth}
      \centering
      \includegraphics[width=1.0\linewidth]{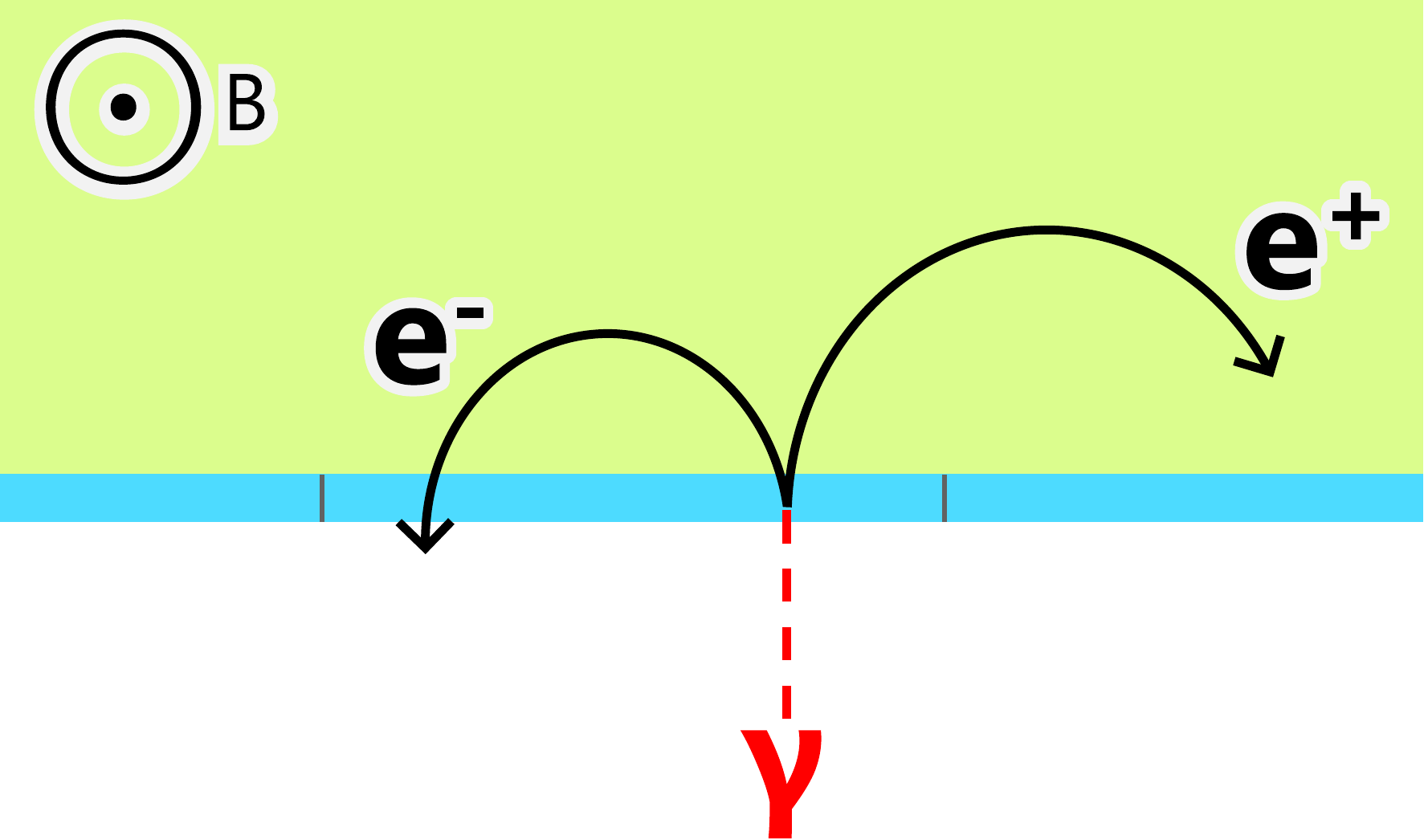}
      \subcaption{ }
      \label{fig:BoomerangHalfTurn}
   \end{minipage}
   \begin{minipage}[t]{0.48\linewidth}
      \centering
      \includegraphics[width=1.0\linewidth]{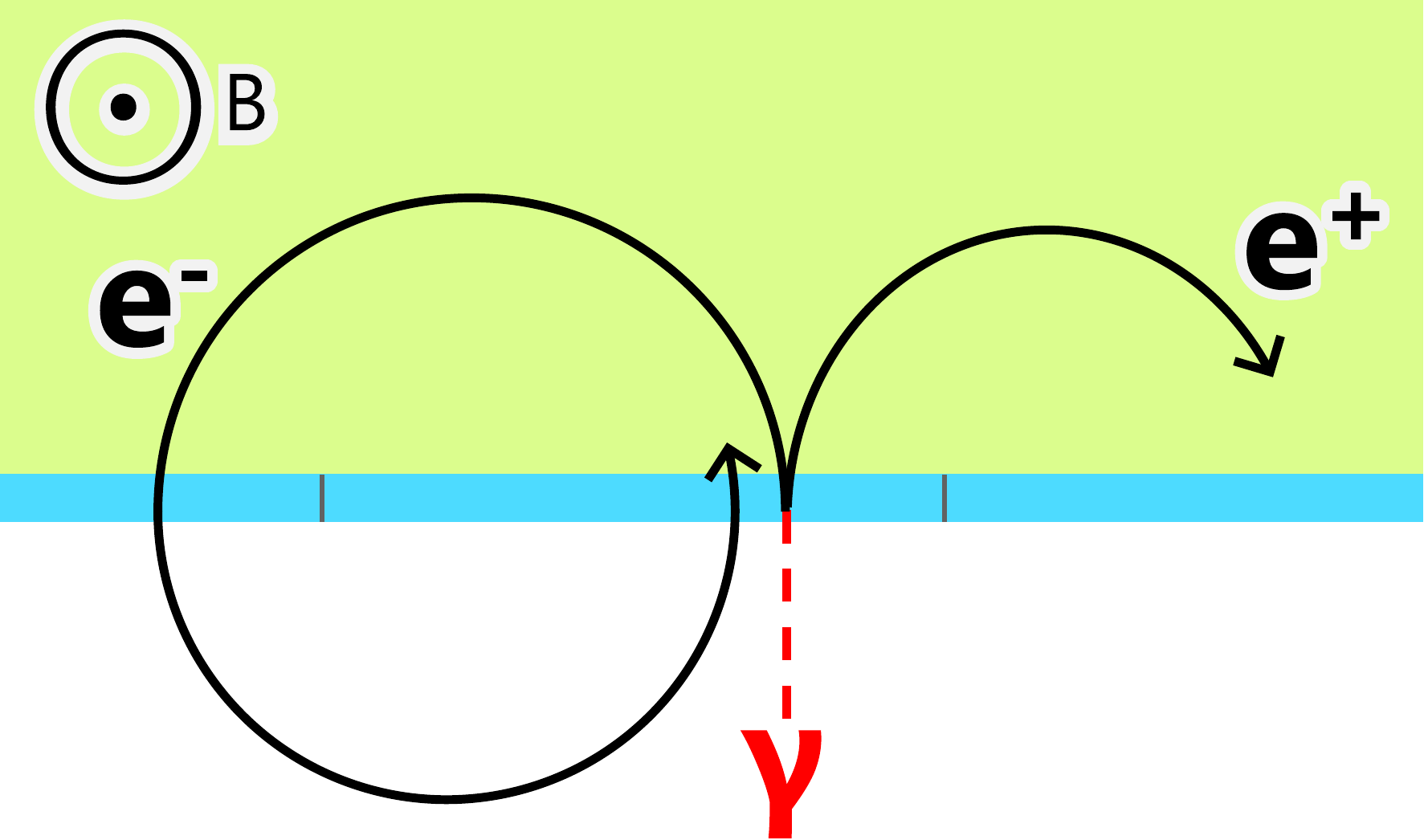}
      \subcaption{ }
      \label{fig:BoomerangMultiTurns}
   \end{minipage}
   \caption{
      Events where a conversion particle re-enters the same cell in which the conversion occurred.
      (a)~After a half turn. This can be identified by the pair-tracker (green shaded region); such events are rejected and do not contribute to the energy spectrum in \figref{fig:EGammaRec}.
      (b)~After more than a full turn, where the re-entering trajectory is outside the tracker volume.
      This contributes to the high-energy tail in \figref{fig:EGammaRec} due to the overestimation of $E_\mathrm{dep}$.
   }
   \label{fig:Boomerang}
\end{figure}

\noindent The impact on $\eff{topo}$ in \eqref{eq:efficiency} is the primary factor to consider when optimizing the segmentation size of the converter.
As discussed in \secref{sec:EfficiencySimulation}, conversion particles may re-enter the same cell in which the conversion occurred after a half turn or more, 
as illustrated in \figref{fig:Boomerang}.
In the case of a half-turn re-entry (\figref{fig:BoomerangHalfTurn}), such events can be identified from the tracks reconstructed by the pair-tracker and subsequently discarded.
In contrast, for multiple-turn re-entries (\figref{fig:BoomerangMultiTurns}), which cannot be identified by the pair-tracker 
\footnote{Depending on the design, some of these may be identified by the pair-tracker coupled to an inner converter layer.}, 
the measured $E_\mathrm{dep}$ in \eqref{eq:EGammaReco} also includes the additional energy deposited at the re-entry points.
This leads to an overestimation of $E_\gamma$, producing a high-energy tail in the reconstructed energy spectrum.
While a low-energy tail primarily reduces the signal efficiency, a high-energy tail directly increases the background contamination in the signal region and degrades the sensitivity of the \meg search.
This is because the spectrum of background photons from radiative decay falls steeply toward the kinematic endpoint, where the signal is expected.
Therefore, suppressing this high-energy tail is particularly important.

\par
\begin{figure}[tbp]
\centering
\begin{minipage}[t]{0.48\linewidth}
    \centering
    \includegraphics[width = 1.\linewidth]{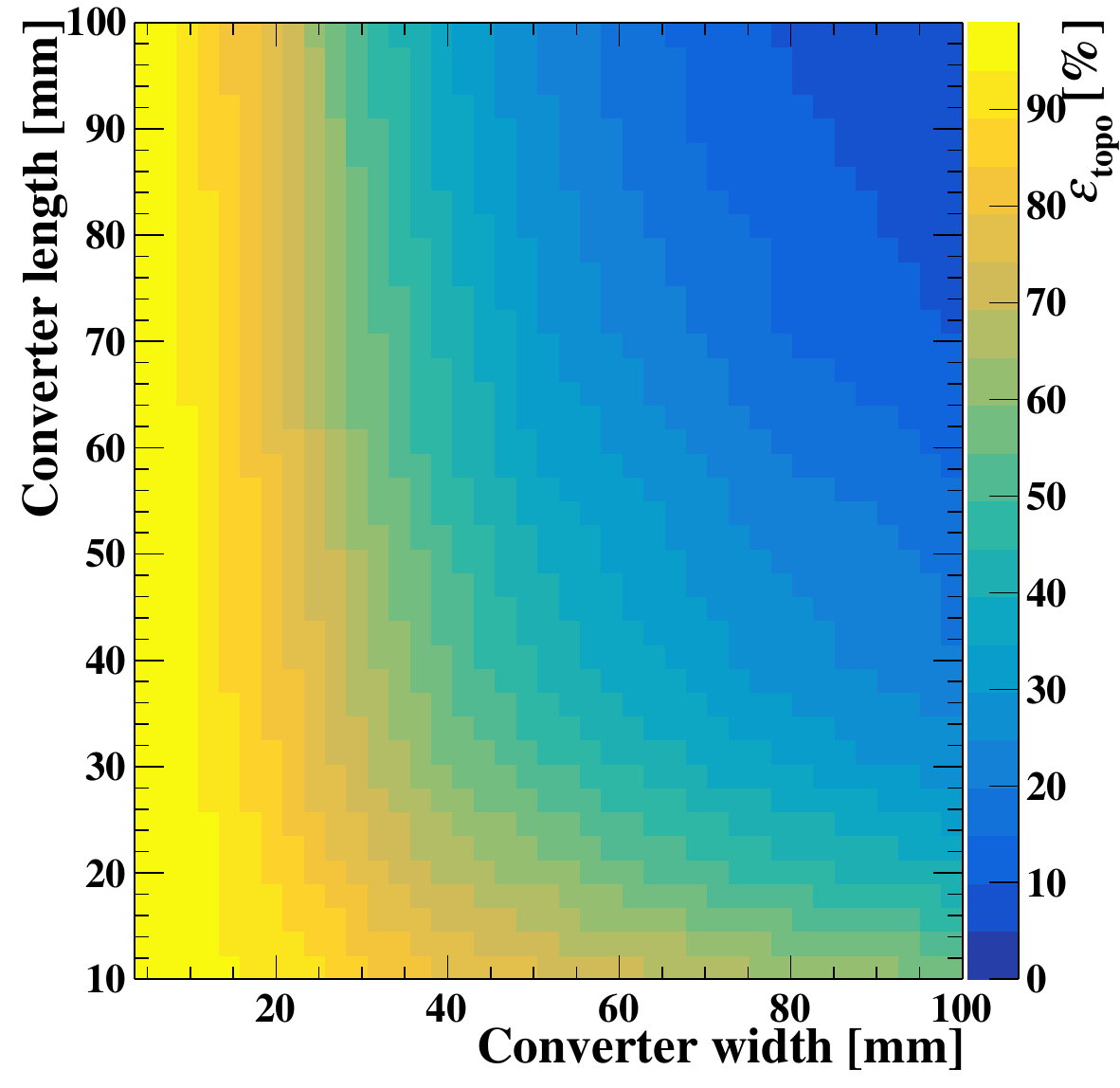}
    \subcaption{$\theta_\gamma = \SI{90}{\degree}$}
    \label{fig:converter_segmentation90deg}
\end{minipage}
\begin{minipage}[t]{0.48\linewidth}
    \centering
    \includegraphics[width = 1.\linewidth]{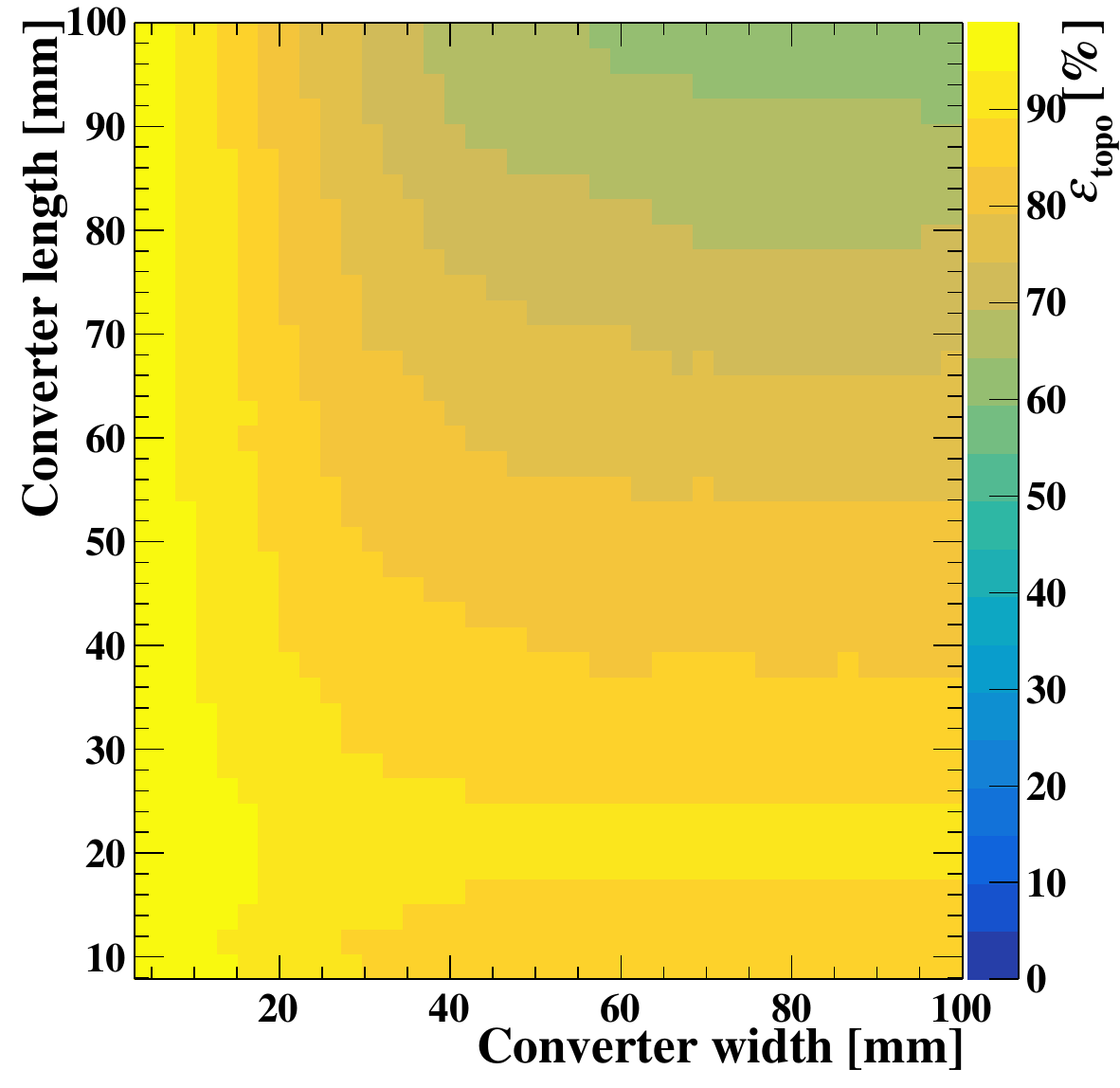}
    \subcaption{$\theta_\gamma = \SI{30}{\degree}$}
    \label{fig:converter_segmentation30deg}
\end{minipage}
\caption{Simulated $\eff{topo}$ value with various converter cell dimensions, at two different signal photon injection angles $\theta_\gamma$. The length (width) is oriented along the $z$ $(\phi)$ direction in \figref{fig:fullsketch}.}
\label{fig:converter_segmentation}
\end{figure}

\indent We performed simulations to investigate the dependence of $\eff{topo}$ on the crystal segmentation along both the $z$ and $\phi$ directions (see \figref{fig:fullsketch} for definitions).
The results are shown in \figref{fig:converter_segmentation}.
The two plots correspond to simulations performed at different incident photon angles, $\theta_\gamma = \ang{90}$ and $\ang{30}$, 
with LYSO thicknesses of \SI{3}{mm} and \SI{1.5}{mm}, respectively.
For $\theta_\gamma = \ang{90}$, the $z$-momentum component is small, so the $e^+e^-$ pairs follow nearly circular trajectories, 
increasing the probability of returning to the same converter cell after a full turn.
Consequently, segmentation in the $\phi$ direction has a stronger impact on $\eff{topo}$ than segmentation in $z$.
In contrast, for $\theta_\gamma = \ang{30}$, the particles have a larger $z$-momentum component, resulting in more elongated trajectories.
They are therefore more likely to re-enter a different $z$-cell, making $\eff{topo}$ more sensitive to $z$ segmentation than to $\phi$ segmentation.
Taking into account the total number of readout channels, the baseline segmentation was chosen to be \SI{5}{mm} in width ($\phi$ direction) and \SI{50}{mm} in length ($z$ direction), which yields an $\eff{topo}$ of $\SI{95.6}{\percent}$.
\par
\begin{figure}[tbp]
   \centering
   \includegraphics[width = 0.7\linewidth]{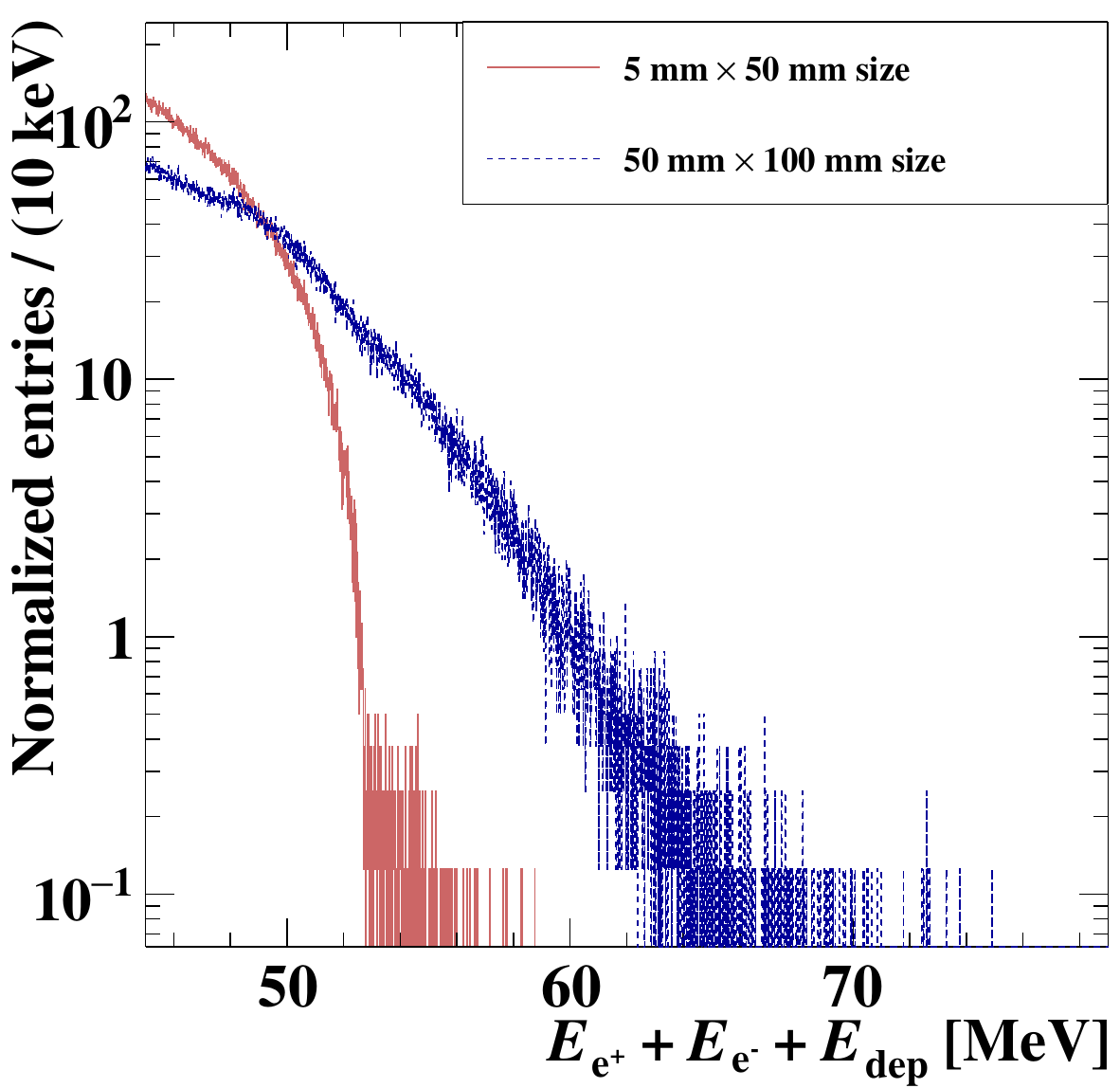}
   \caption{
      Simulated $E_\gamma$ spectra for photons from the radiative decay of muons, compared between two different converter segmentation sizes. }
   \label{fig:RMDspectrum}
\end{figure}

%%%%%%%%%%%%%%%%%%%%%%%%%%
%% BG high energy tail  %%
%%%%%%%%%%%%%%%%%%%%%%%%%%
\indent A second consideration regarding the crystal segmentation is its impact on the background in the \meg search.
The high-energy tail observed in the signal $E_\gamma$ spectrum, induced by the overestimated $E_\mathrm{dep}$, 
also appears in the background spectrum, which is another important factor affecting the \meg search sensitivity.
To study this effect, the energy reconstruction was simulated using the dominant source of background photons: radiative muon decay ($\mu \to e \nu \bar{\nu} \gamma$).
\figref{fig:RMDspectrum} shows the simulated background spectra scaled to \num{1e15} muon decays for two different converter segmentations.
Insufficient segmentation enhances the high-energy tail above \SI{52.8}{MeV}, thereby increasing the number of background events in the \meg search
and degrading the sensitivity.
This effect, however, was found not to be critical with the $\SI{5}{mm} \times \SI{50}{mm}$ segmentation: 
only 34 events are reconstructed above \SI{52.8}{MeV}, compared to 4452 events for the much coarser segmentation shown in \figref{fig:RMDspectrum}.
This significant suppression of the high-energy tail further supports the choice of the baseline design.
\par
\begin{figure}[tbp]
   \centering
   \includegraphics[width = 0.7\linewidth]{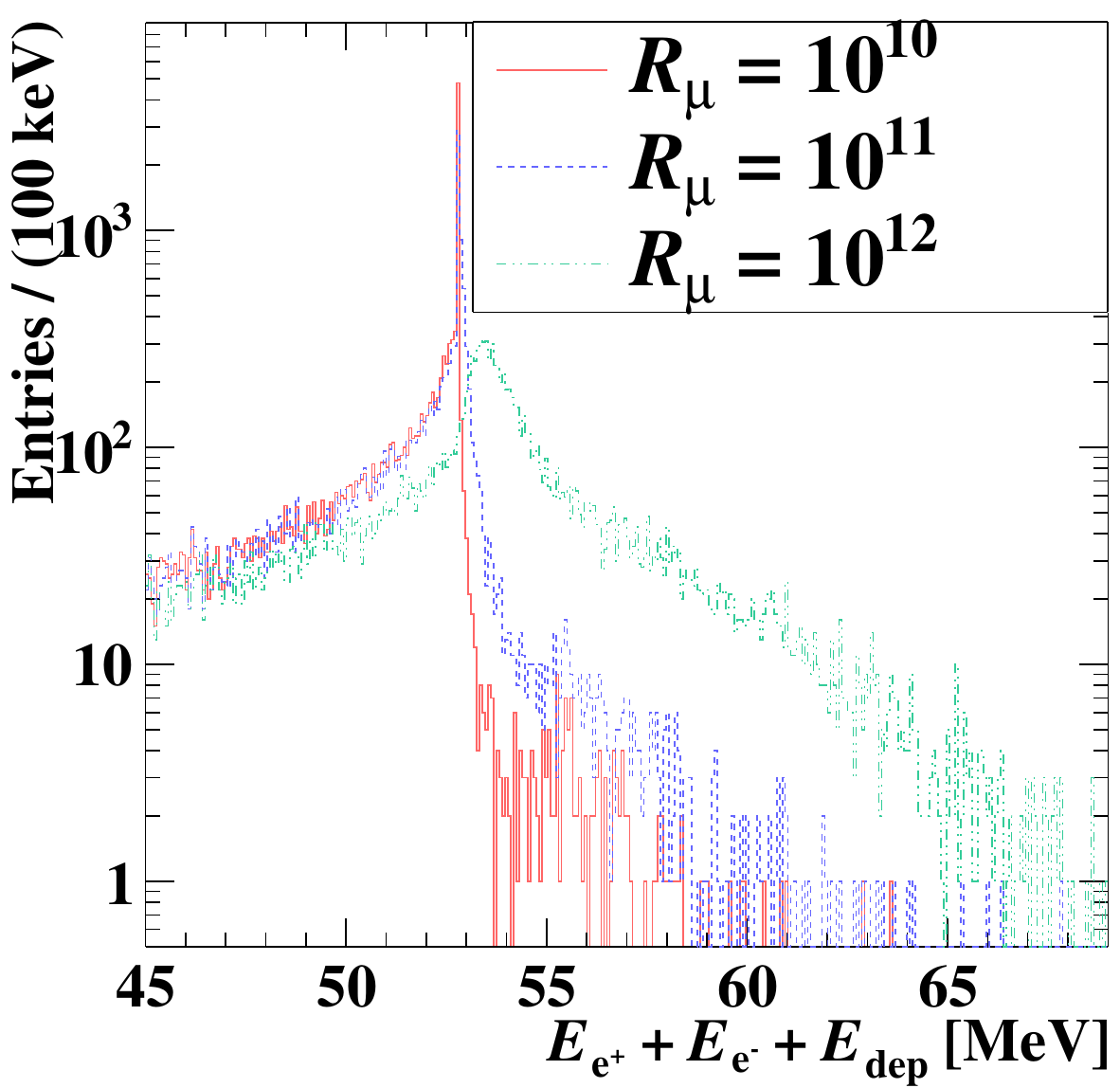}
   \caption{
      $E_\gamma$ spectra in signal photon measurement simulated with pile-up effects at several assumed muon beam rates.  
      Photons from the radiative decay of muons are considered as the source of pile-up. 
   }
   \label{fig:pileup_effect}
\end{figure}

%%%%%%%%%%%%%%%%%%%%%%%%%%
%% Rate capability      %%
%%%%%%%%%%%%%%%%%%%%%%%%%%
\indent Our final consideration regarding segmentation is the rate capability, which is determined by the pile-up of background photons detected within the same converter cell.
In such pile-up events, $E_\mathrm{dep}$ is overestimated due to the additional energy deposited by the pile-up photons, 
creating another source for the high-energy tail in the reconstructed energy spectrum.
The impact of this effect is governed by the probability of such coincidences, which depends on the geometrical acceptance of the crystal cell and the muon rate.
In this study, we simulated the impact on the signal energy spectrum at several beam rates for a crystal located \SI{20}{cm} away from the beam axis with the 
baseline segmentation of $\SI{5}{mm} \times \SI{50}{mm}$, as shown in \figref{fig:pileup_effect}.
Here, we focused on the pile-up effect at $\theta_\gamma = \ang{90}$, where the impact is expected to be strongest due to the large geometrical acceptance of the cell.
The pile-up photons were generated from the radiative muon decay spectrum, and their energy depositions within a \SI{200}{ns} coincidence window were added to $E_\mathrm{dep}$.
With the baseline converter segmentation, a distortion of the spectrum appears only with $R_\mu$ above \SI{e11}{\mu^+\per\second}, 
which is beyond the expected range of the HIMB project.
This result also supports the choice of \SI{5}{mm}-wide and \SI{50}{mm}-long segmentation as a feasible design.

%%%%%%%%%%%%%%%%%%%%%%%%%%
%% Light yield & E_dep  %%
%%%%%%%%%%%%%%%%%%%%%%%%%%
\subsection{Requirement on active converter light yield}\label{sec:ConverterLYrequirement}
\begin{figure}[tbp]
   \centering
   \includegraphics[width=0.7\linewidth]{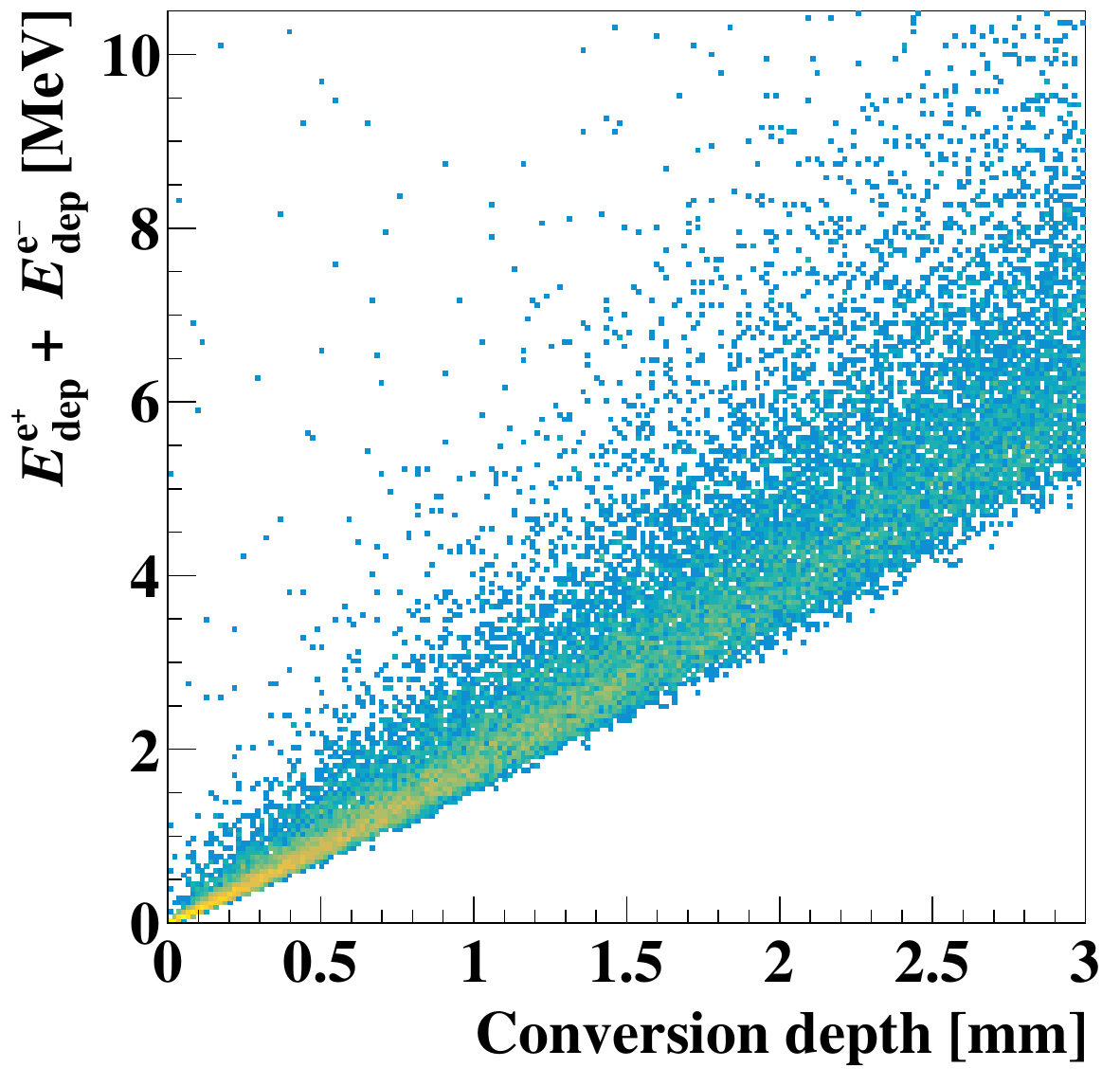}
   \caption{
      Scatter plot of the total energy deposit inside the converter vs. the conversion depth, namely $x$ defined in \eqref{eq:EfficiencyPhysicalProcess}.
   }
   \label{fig:DepthvsEdep}
\end{figure}

\noindent The energy deposited by an $e^+e^-$ pair inside the \SI{3}{mm}-thick converter ranges from nearly zero up to 
approximately \SI{10}{MeV}, depending on the conversion depth, as shown in \figref{fig:DepthvsEdep}.
The target energy resolution for the measurement of a $\SI{52.8}{MeV}$ photon is \SI{200}{keV} (\SI{0.4}{\percent}), as stated in \secref{sec:Introduction}.
This requirement translates into a precision of \SI{2}{\percent} for the maximum energy deposit of $E_\mathrm{dep} = \SI{10}{MeV}$.
To ensure that statistical fluctuations in the number of photoelectrons are suppressed to a sufficiently low level, the following condition must be satisfied:
\begin{equation}
   \frac{1}{\sqrt{N_\mathrm{p.e.}}} \leq \SI{2}{\percent}.
\end{equation}
Therefore, the active converter is required to have a light yield of at least 2500 photoelectrons for a \SI{10}{MeV} energy deposit.

%%%%%%%%%%%%%%%%%%%%%%%%%%
%%                      %%
%% Beam test            %%
%%                      %%
%%%%%%%%%%%%%%%%%%%%%%%%%%
\section{Measurement of LYSO performance with an electron beam}\label{sec:BeamTest}
\begin{figure}[tbp]
   \centering
   \includegraphics[width=0.7\linewidth]{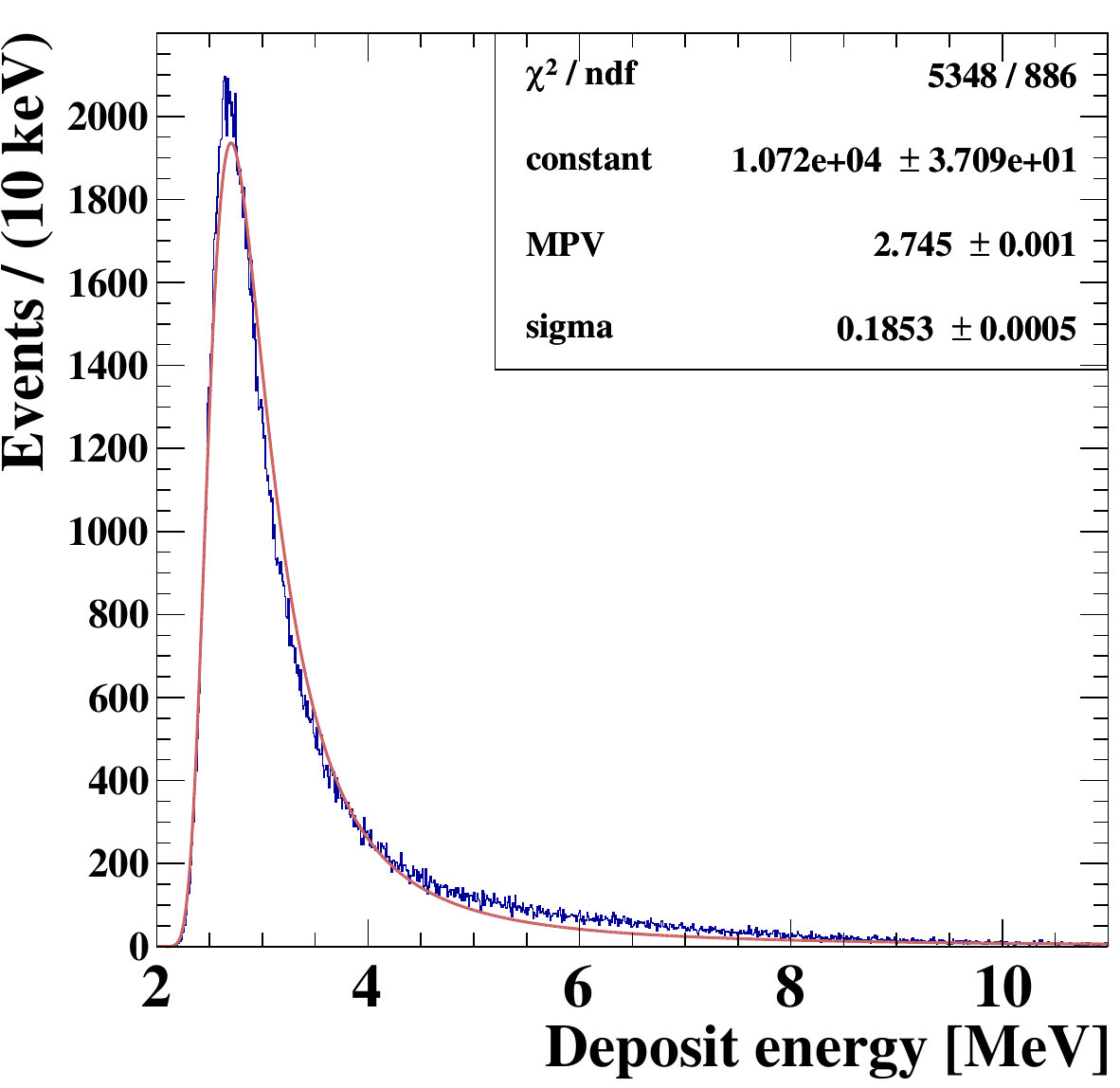}
   \caption{
      Simulated energy deposit inside a \SI{3}{mm}-thick LYSO for a perpendicularly injected \SI{3}{GeV} electron beam. 
	   A Landau fit is overlaid as the red curve.
   }
   \label{fig:EdepBeamtest}
\end{figure}

\noindent We conducted a beam test using \SI{3}{GeV} electrons to evaluate whether the LYSO-based active converter satisfies the resolution requirements for the future \meg experiment.
To achieve the target resolutions of $\Delta t_\gamma < \SI{30}{ps}$ and $\Delta E_\gamma < \SI{200}{keV}$, the active converter must provide a time resolution of approximately \SI{40}{ps} for a single conversion particle and a light yield of at least 2500 photoelectrons for an energy deposit of \SI{10}{MeV}, as discussed in \secref{sec:DesignConcept} and \secref{sec:ConverterLYrequirement}.
The simulated energy deposit of a \SI{3}{GeV} electron is \SI{2.7}{MeV} at the most probable value (MPV) of the Landau distribution (\figref{fig:EdepBeamtest}). 
We therefore aimed to confirm a light yield exceeding 700 photoelectrons in this beam test, which scales to the requirement at \SI{10}{MeV}.
Regarding the timing performance, achieving the \SI{40}{ps} requirement with the \SI{3}{GeV} electron beam is considered sufficient.
This is because such a particle deposits an amount of energy comparable to, or smaller than, that of a typical conversion particle re-entering the converter after being bent by the magnetic field.
Therefore, demonstrating a time resolution of \SI{40}{ps} at \SI{3}{GeV} ensures adequate performance under actual operating conditions.
%%%%%%%%%%%%%%%%%%%%%%%%%%
%%                      %%
%% Experimental setup   %%
%%                      %%
%%%%%%%%%%%%%%%%%%%%%%%%%%
\subsection{Experimental setup}
\subsubsection{Active converter prototypes}\label{sec:ActiveConverterPrototype}
\noindent The {\prototype}s consist of Ce-doped LYSO crystals (provided by JT Crystal Technology, JTC) wrapped in ESR reflective film.
JTC provides two types of LYSO: ``Ce:FTRL'' (fast response time, moderate light yield) and ``Ce:LYSO'' (high light yield, moderate response time), 
whose properties are summarized in \tabref{tab:LYSOproperties}.
To maximize timing performance, Ce:FTRL was selected for the {\prototype}s.
Based on the optimization studies in \secref{sec:Simulation}, we primarily tested LYSO crystals with dimensions of \lysodimension{50}{5}{3}.
In addition, a \SI{1.5}{mm}-thick crystal was evaluated for potential use in the outer regions of the detector, as discussed in \secref{sec:Simulation}.

\begin{table}[tbp]
    \centering
    \begin{tabular}{ccccc}\hline
         Properties                       & Ce:FTRL               & Ce:LYSO \\\hline\hline
         \small \begin{tabular}{c} Coincident time resolution \\with $\SI{2}{mm}$ cube (ps) \end{tabular}
         & 96 & 125 \\\hline
         \small Light output (photons / MeV)     
         & $30000\pm\SI{10}{\percent}$ & $36000\pm\SI{10}{\percent}$ \\\hline
         \small Decay time (ns)                  
         & 31 & 40    \\\hline
         \small \begin{tabular}{c} Wavelength of \\ max emission(nm)\end{tabular}
         & 420& 420   \\\hline
         \small refractive index                 
         & 1.81 & 1.81  \\\hline
         \small density(\SI{}{g/cm^3})                     
         & 7.2 & 7.2   \\  \hline
    \end{tabular}
    \caption{Properties of fast-type LYSO (Ce:FTRL) and the normal-type LYSO (Ce:LYSO), provided by JTC.}
    \label{tab:LYSOproperties}
\end{table}
\begin{table}[tbp]
    \centering
    \begin{tabular}{cccc}\hline
        \footnotesize SiPM model             
        & \footnotesize\begin{tabular}{c}photosensitive\\ area ($\SI{}{mm^2}$)\end{tabular}        
        & \footnotesize\begin{tabular}{c}SiPMs\\ per side\end{tabular} 
        & \footnotesize overvoltage\\\hline\hline
        \footnotesize S14160-3050HS         
        &\footnotesize $3\times 3$ &\footnotesize 3 &\footnotesize $+ \SI{6}{V}$            \\\hline
        \footnotesize S14160-6050HS         
        &\footnotesize $6\times 6$ &\footnotesize 1 &\footnotesize $+ \SI{5}{V}$             \\\hline
        \footnotesize \begin{tabular}{c}MICROFJ-40035\\-TSV-TR1\end{tabular} 
        &\footnotesize $4\times 4$ &\footnotesize 1 &\footnotesize $+ \SI{5}{V}$             \\\hline
    \end{tabular}
    \caption{SiPM types and their number of channels per LYSO readout face, which were used in the active converter prototypes.}
    \label{tab:SiPM}
\end{table}
\indent Scintillation light was collected from both ends of the LYSO bar in all configurations to ensure high light collection efficiency. To clarify the sensor geometry and readout topology, we categorize the configurations into three types based on the number of SiPMs and their electrical connection:
\begin{description}
\item[Double-sided Single-SiPM:] One SiPM was mounted on each end of the crystal (2 channels in total). This configuration was used for SiPM models with a large photosensitive area (e.g., S14160-6050HS and MICROFJ-40035).
\item[Double-sided Series-connected:] Three $3\times \SI{3}{mm^2}$ SiPMs (S14160-3050HS) were mounted on each end and electrically connected in series. This provides a single summed signal per side (2 channels in total).
\item[Double-sided Individual:] Three $3\times \SI{3}{mm^2}$ SiPMs (S14160-3050HS) were mounted on each end, with all six SiPMs read out independently (6 channels in total).
\end{description}

\newcounter{rowno}
\renewcommand{\therowno}{\Roman{rowno}}
\newcommand{\rowlabel}[1]{
  \refstepcounter{rowno}
  (\therowno)\label{#1}%
}
\newcommand{\rowref}[1]{(\ref{#1})}
\begin{table}[tbp]
   \centering\small
   \begin{tabular}{cccc}\hline
      No. & LYSO thickness & SiPM model    & Readout scheme  \\ \hline\hline
      \rowcolor{black!30}\multicolumn{4}{l}{2024 Beam Test}  \\ \hline
      \rowlabel{row:20243mmSiPM}         & \SI{3}{mm}     & S14160-3050HS & Series-connected \\ \hline
      \rowlabel{row:20246mmSiPM}         & \SI{3}{mm}     & S14160-6050HS & Single-SiPM      \\ \hline
      \rowlabel{row:20244mmSiPM}         & \SI{3}{mm}     & MICROFJ-40035 & Single-SiPM      \\ \hline
      \rowlabel{row:20246mmSiPMThinLYSO} & \SI{1.5}{mm}   & S14160-6050HS & Single-SiPM      \\ \hline
      \rowcolor{black!30}\multicolumn{4}{l}{2023 Beam Test}\\ \hline
      \rowlabel{row:2023Series}          & \SI{3}{mm}     & S14160-3050HS & Series-connected \\ \hline
      \rowlabel{row:2023Independent}     & \SI{3}{mm}     & S14160-3050HS & Individual       \\ \hline
   \end{tabular}
   \caption{Summary of the measurement configurations. The readout schemes are categorized based on the number of sensors and their electrical connection: Single-SiPM'' (one sensor per side), Series-connected'' (three sensors summed), and ``Individual'' (three sensors read out separately).}
   \label{tab:Configurations}
\end{table}
\indent Throughout this series of tests, the performance of the prototypes was evaluated in various configurations, combining different crystal thicknesses, SiPM models, and readout schemes.
The configurations discussed in this paper are summarized in \tabref{tab:Configurations}, along with the year in which each beam test was conducted.

\subsubsection{Readout electronics}\label{sec:ReadOutElectronics}
\begin{figure}[tbp]
    \centering
    \includegraphics[width=1\linewidth]{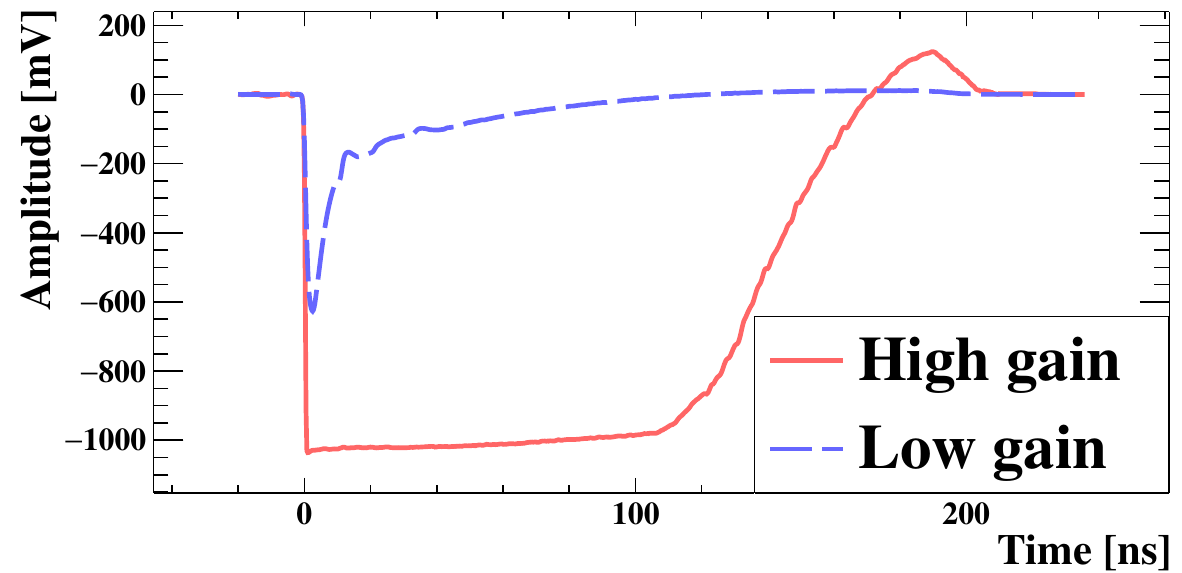}
    \caption{
       Average waveforms in the low- and high-gain channels of the \prototype.
       The high-gain waveform is clipped at $-\SI{1000}{mV}$ because of the dynamic range of the DRS4 digitizer.}
    \label{fig:waveform}
\end{figure}

\noindent Triggering and data acquisition were performed using a WaveDREAM board~\cite{WDB}, which integrates two DRS4 waveform digitizer chips~\cite{RITT2004470} along with built-in amplification and shaping circuits.
The detection of the first-arriving scintillation photons, characterized by a steep rising edge and minimal timing jitter, is crucial for precise time measurements.
Consequently, a high amplifier gain is desirable for timing purposes.
In contrast, accurate charge measurement requires the waveform amplitude to remain within the dynamic range of the digitizer.
To satisfy both requirements, signals were split into two branches and recorded with different amplifier gain settings (``high gain'' and ``low gain'') during the 2024 beam test campaign, at a sampling frequency of \SIrange{4}{5}{GSPS}.
In the 2023 beam test, which primarily focused on evaluating the timing performance, signal waveforms were recorded using only the high-gain setting.
Typical waveforms obtained from the high- and low-gain channels are shown in \figref{fig:waveform}.

\subsubsection{Data-taking scheme}
\noindent Beam test campaigns were conducted at the KEK PF-AR Test Beam Line (Tsukuba campus) in December 2023 and 2024.
The beam line delivered electrons with a momentum of approximately \SI{3}{GeV/c} at a rate of about \SI{4.5}{kHz}.
\figref{fig:setup} shows the overall experimental configuration.
Two LYSO converter prototypes were installed simultaneously during the tests.
This setup allowed for efficient use of the available beam time and enabled the evaluation of timing performance using three independent measurements from the reference counter, the upstream prototype, and the downstream prototype, as defined in \eqref{eq:timeResolutionEvaluation}.
To investigate the dependence of the performance on the beam incidence position and angle, the prototypes were mounted on a movable stage (\figref{fig:positionScan_setup}) and a rotating stage (\figref{fig:angleScan_setup}), respectively.
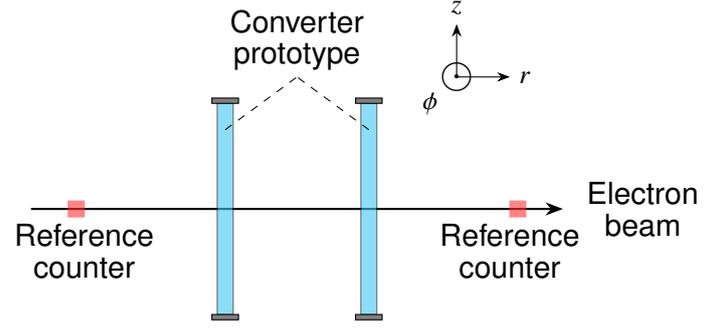
\begin{figure}[tbp]
   \hspace*{-25pt}
    \begin{tikzpicture}[scale=0.7, font=\sffamily]
      \draw[thick, -Stealth] (-5,0) -- (5,0);
      \filldraw[fill=cyan!60, opacity=0.6] (-1.2,-2) -- ++(0,4) -- ++(-0.3,0) -- ++(0,-4) -- cycle; % DS LYSO
      \filldraw[fill=cyan!60, opacity=0.6] (1.2,-2) -- ++(0,4) -- ++(0.3,0) -- ++(0,-4) -- cycle; % DS LYSO

      \filldraw[fill=gray] (-1.1, -2) -- ++(0,-0.1) -- ++ (-0.5, 0) -- ++ (0,0.1) -- cycle; %DS SiPM down
      \filldraw[fill=gray] (-1.1, +2) -- ++(0,+0.1) -- ++ (-0.5, 0) -- ++ (0,-0.1) -- cycle;%DS SiPM up
      \filldraw[fill=gray] (1.1, -2) -- ++(0,-0.1) -- ++ (0.5, 0) -- ++ (0,0.1) -- cycle;   %US SiPM down
      \filldraw[fill=gray] (1.1, +2) -- ++(0,+0.1) -- ++ (0.5, 0) -- ++ (0,-0.1) -- cycle;  %US SiPM up

      \filldraw[red!80, opacity=0.6] (-4,-0.15) -- ++(-0.3,0) -- ++ (0,0.3) -- ++(0.3,0) --cycle;
      \filldraw[red!80, opacity=0.6] (4,-0.15) -- ++(0.3,0) -- ++ (0,0.3) -- ++(-0.3,0) --cycle;

      \draw[dashed] (-1.35, 1.5) -- (0,2.5);
      \draw[dashed] (1.35, 1.5) -- (0,2.5);

      \node[align=center, text width = 3cm] at (0,3.2) {\large Converter\\prototype};
      % \node[align=center, text width = 3cm] at (-1.5,2.8) {\large Converter\\prototype};
      \node[align=center, text width = 4cm] at (4,-0.8)  {\large Reference\\counter};
      \node[align=center, text width = 4cm] at (-4,-0.8)  {\large Reference\\counter};
      \node[align=center, text width = 4cm] at (6.5,0)  {\large Electron\\beam};

      % \draw[-Stealth] (3,2.5) -- ++(1,0) node [right] {$r$};
      % \draw[-Stealth] (3,2.5) -- ++(0,1) node [above] {$z$};
      % \node at (3,2.5) {$\bigodot$};
      % \node at (2.5,2) {$\phi$};
      
\end{tikzpicture}
    \caption{
    Schematic of the electron beam test setup.
    % The coordinate system follows the convention defined in \figref{fig:fullsketch}.
    }
    \label{fig:setup}
\end{figure}
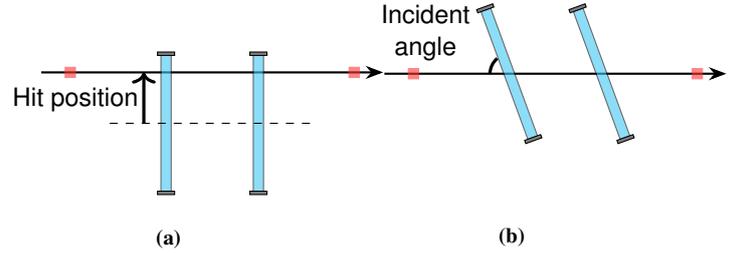
\begin{figure}[tbp]
   \begin{minipage}{0.48\linewidth}
      \begin{tikzpicture}[scale=0.45, font=\sffamily,baseline={(0,0)}]
   \useasboundingbox (-6, -4) rectangle (6,2); 
      \draw[thick, -Stealth] (-5,0) -- (5,0);

      \filldraw[fill=cyan!60, opacity=0.6] (-1.2,-3.5) -- ++(0,4) -- ++(-0.3,0) -- ++(0,-4) -- cycle; % DS LYSO
      \filldraw[fill=gray] (-1.1, -3.5) -- ++(0,-0.1) -- ++ (-0.5, 0) -- ++ (0,0.1) -- cycle; %DS SiPM down
      \filldraw[fill=gray] (-1.1, +0.5) -- ++(0,+0.1) -- ++ (-0.5, 0) -- ++ (0,-0.1) -- cycle;%DS SiPM up
      \filldraw[fill=cyan!60, opacity=0.6] (1.2,-3.5) -- ++(0,4) -- ++(0.3,0) -- ++(0,-4) -- cycle; % US LYSO
      \filldraw[fill=gray] (1.1, -3.5) -- ++(0,-0.1) -- ++ (0.5, 0) -- ++ (0,0.1) -- cycle;   %US SiPM down
      \filldraw[fill=gray] (1.1, +0.5) -- ++(0,+0.1) -- ++ (0.5, 0) -- ++ (0,-0.1) -- cycle;  %US SiPM up

      \draw[dashed] (-3,-1.5) -- (3,-1.5);
      \draw[->, very thick] (-2,-1.5) -- ++(0,1.5);
      % \node[align=center, text width = 4cm]  at (-3.3,-0.8) {\normalsize Hit \\ position};
      \node at (-4,-0.8) {\normalsize Hit position};

      \filldraw[red!80, opacity=0.6] (-4,-0.15) -- ++(-0.3,0) -- ++ (0,0.3) -- ++(0.3,0) --cycle;
      \filldraw[red!80, opacity=0.6] (4,-0.15) -- ++(0.3,0) -- ++ (0,0.3) -- ++(-0.3,0) --cycle;

\end{tikzpicture}
      \subcaption{}
      \label{fig:positionScan_setup}
   \end{minipage}
   \hspace{0.01\linewidth}
   \begin{minipage}{0.48\linewidth}
      \begin{tikzpicture}[scale=0.45, font=\sffamily,baseline={(0,0)}]
   \useasboundingbox (-6, -4) rectangle (6,2); 
      \draw[thick, -Stealth] (-5,0) -- (5,0);

      \begin{scope}[rotate around={20:(-1.35,0)}]
         \filldraw[fill=cyan!60, opacity=0.6] (-1.2,-2) -- ++(0,4) -- ++(-0.3,0) -- ++(0,-4) -- cycle; % DS LYSO
         \filldraw[fill=gray] (-1.1, -2) -- ++(0,-0.1) -- ++ (-0.5, 0) -- ++ (0,0.1) -- cycle; %DS SiPM down
         \filldraw[fill=gray] (-1.1, +2) -- ++(0,+0.1) -- ++ (-0.5, 0) -- ++ (0,-0.1) -- cycle;%DS SiPM up
      \end{scope}
      \begin{scope}[rotate around={20:(1.35,0)}]
         \filldraw[fill=cyan!60, opacity=0.6] (1.2,-2) -- ++(0,4) -- ++(0.3,0) -- ++(0,-4) -- cycle; % US LYSO
         \filldraw[fill=gray] (1.1, -2) -- ++(0,-0.1) -- ++ (0.5, 0) -- ++ (0,0.1) -- cycle;   %US SiPM down
         \filldraw[fill=gray] (1.1, +2) -- ++(0,+0.1) -- ++ (0.5, 0) -- ++ (0,-0.1) -- cycle;  %US SiPM up
      \end{scope}
      \draw[very thick] (-1.7, 0.45) arc[start angle=130, end angle=180, radius=0.6];
      \node[align=center, text width = 4cm] at (-3.8, 1.2)  {\normalsize Incident\\angle};

      \filldraw[red!80, opacity=0.6] (-4,-0.15) -- ++(-0.3,0) -- ++ (0,0.3) -- ++(0.3,0) --cycle;
      \filldraw[red!80, opacity=0.6] (4,-0.15) -- ++(0.3,0) -- ++ (0,0.3) -- ++(-0.3,0) --cycle;
\end{tikzpicture}
      \subcaption{}
      \label{fig:angleScan_setup}
   \end{minipage}
   \caption{Schematics of the setups for the (a) beam hit position scan and (b) beam incident angle scan.  }
   \label{fig:Setup_scan}
\end{figure}
\par
\indent Events were triggered by the coincidence of two reference counters placed at the upstream and downstream ends of the setup.
Each counter consisted of a $\SI{5}{mm}$ cubic plastic scintillator read out by a silicon photomultiplier (SiPM), achieving time resolutions of \SIrange{30}{40}{ps}.

\subsection{Data analysis}
\subsubsection{Waveform analysis}
\noindent The waveforms from the {\prototype}s and the reference counters were analyzed to extract several parameters: 
pulse charge, pulse height, and leading-edge timing.
For each waveform, the baseline was first estimated from the pre-pulse region.
The pulse height was then defined as the difference between the peak voltage and the baseline.
The pulse charge was calculated by integrating the baseline-subtracted signal between the start time $t_\mathrm{start}$ and the end time $t_\mathrm{end}$, 
where $t_\mathrm{start}$ and $t_\mathrm{end}$ were identified as the nearest baseline-crossing points before and after the peak time $t_\mathrm{peak}$, respectively.
Finally, the leading-edge timing and time-over-threshold (ToT) were determined using several threshold levels.

\subsubsection{Event selection}\label{sec:EventSelection}
\noindent To evaluate the performance of the {\prototype}s for single MIPs traversing them at controlled incident positions and angles, 
an event selection was applied.
A coincidence of signals in both reference counters was required to ensure that the selected particles from the spatially spread beam passed through the targeted 
region of the LYSO counters.
\par
\begin{figure}[tbp]
   \centering
   \includegraphics[width=0.7\linewidth]{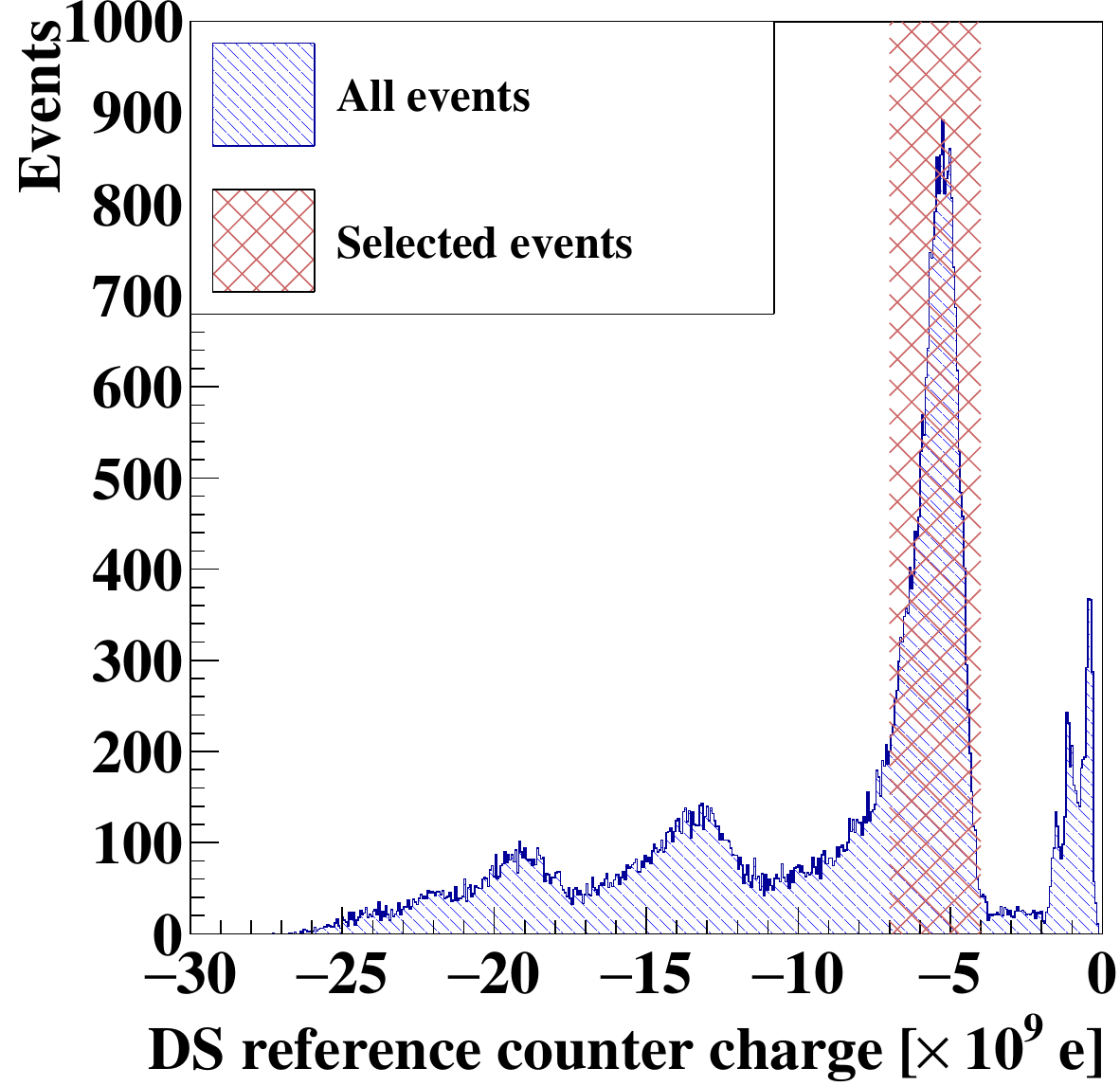}
   \caption{
      Charge distribution of the downstream reference counter.
      % in the 2023 beam test
      Events with charges within the blue shaded region were selected.
   }
   \label{fig:chargeDistDSref_2023}
\end{figure}

\indent In some cases, an incident electron induced an electromagnetic shower in the upstream \prototype, 
resulting in multiple particles entering the downstream one.
Since the analysis targets the response to a single MIP signal, such events must be removed.
These multi-particle events were identified and rejected using the charge distribution of the downstream reference counter.
\figref{fig:chargeDistDSref_2023} shows an example of this distribution: multiple peaks corresponding to several MIPs are visible, and only events consistent with a single MIP
---indicated by the blue shaded region---were retained.
\par
\begin{figure}[tbp]
   \centering
   \includegraphics[width=0.7\linewidth]{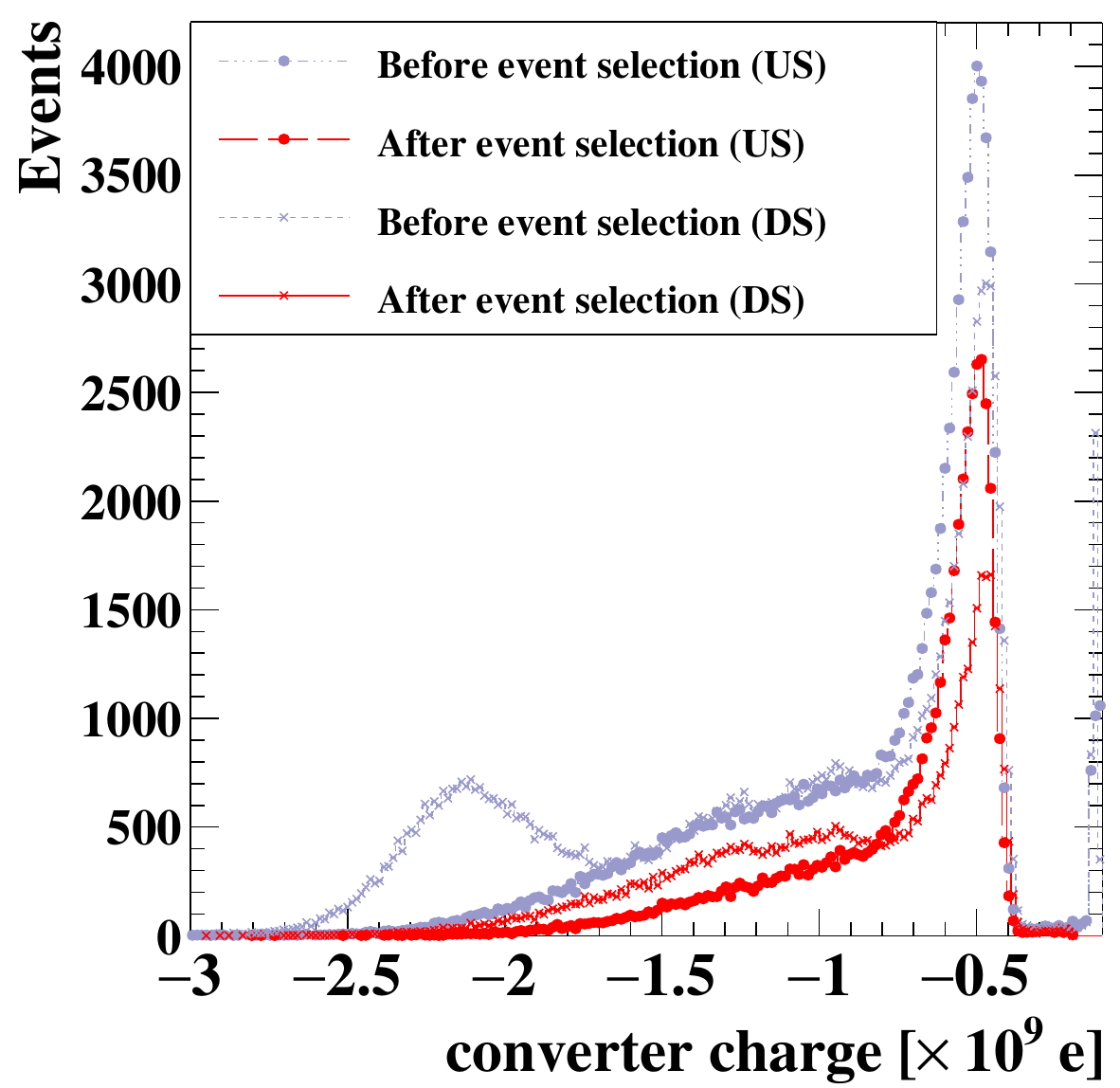}
   \caption{
      Charge distribution of the upstream (US) and downstream (DS) {\prototype}s before and after the event selection.
   }
   \label{fig:eventSelection}
\end{figure}

\indent The impact of this event selection on the 2024 beam-test data is illustrated in \figref{fig:eventSelection}.
The high-charge excess due to multi-particle events is more pronounced in the downstream \prototype~than in the upstream one, 
indicating that most such events originate from showering in the upstream \prototype.
Although this excess is largely suppressed by the selection, a small residual remains in the downstream \prototype.

%%%%%%%%%%%%%%%%%%%%%%%%%%
%%                      %%
%% Timing resolution    %%
%%                      %%
%%%%%%%%%%%%%%%%%%%%%%%%%%
\section{Timing performance}\label{sec:TimingPerformance}
\subsection{Calibration}
\begin{figure}[tbp]
   \centering
   \begin{minipage}[t]{0.48\linewidth}
      \centering
      \includegraphics[width = 1\linewidth]{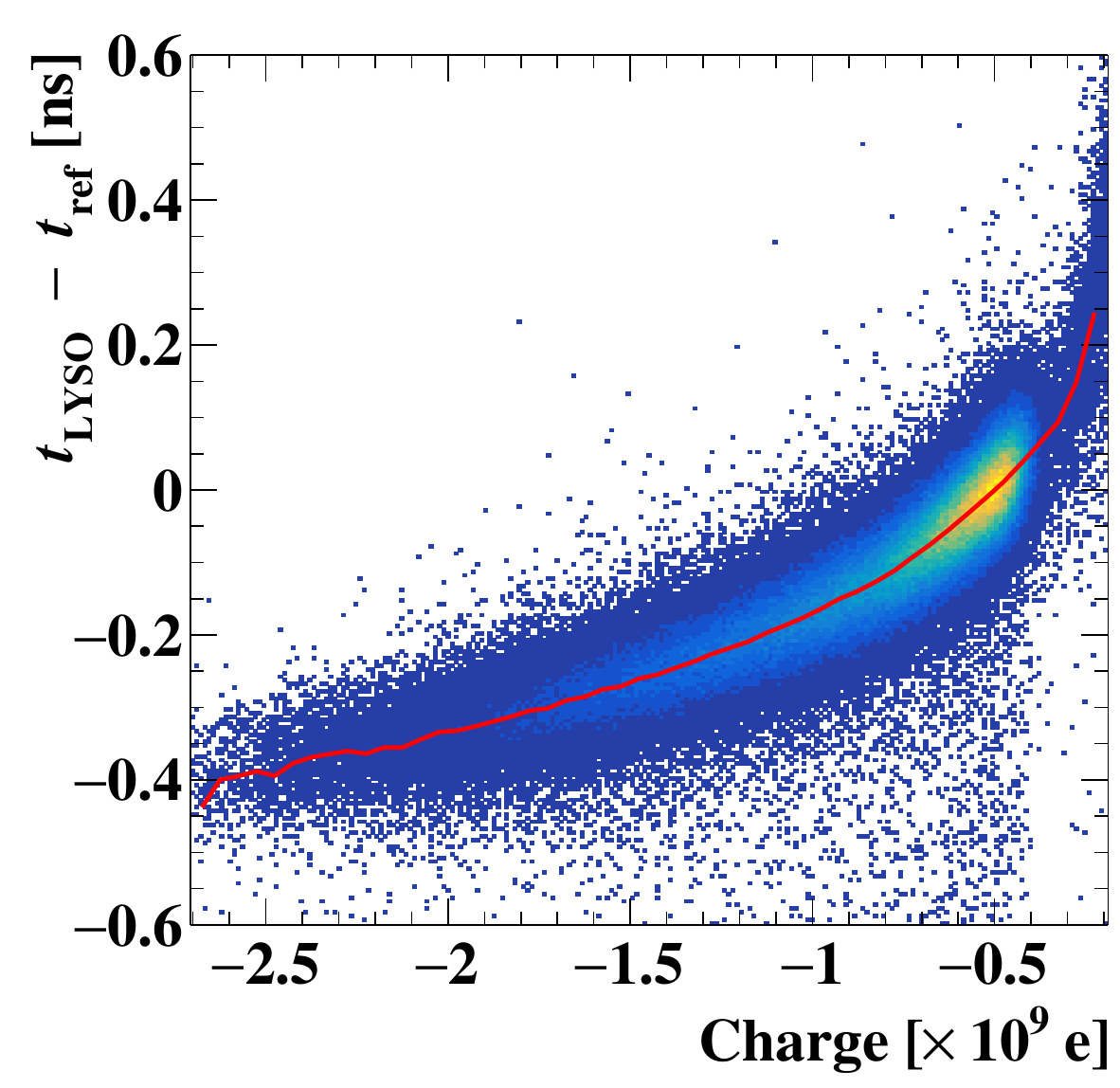}
      \subcaption{2024 data} % 3mm SiPM
      \label{fig:ChargeTimeCorrelation}
   \end{minipage}
   \begin{minipage}[t]{0.48\linewidth}
      \centering
      \includegraphics[width = 1\linewidth]{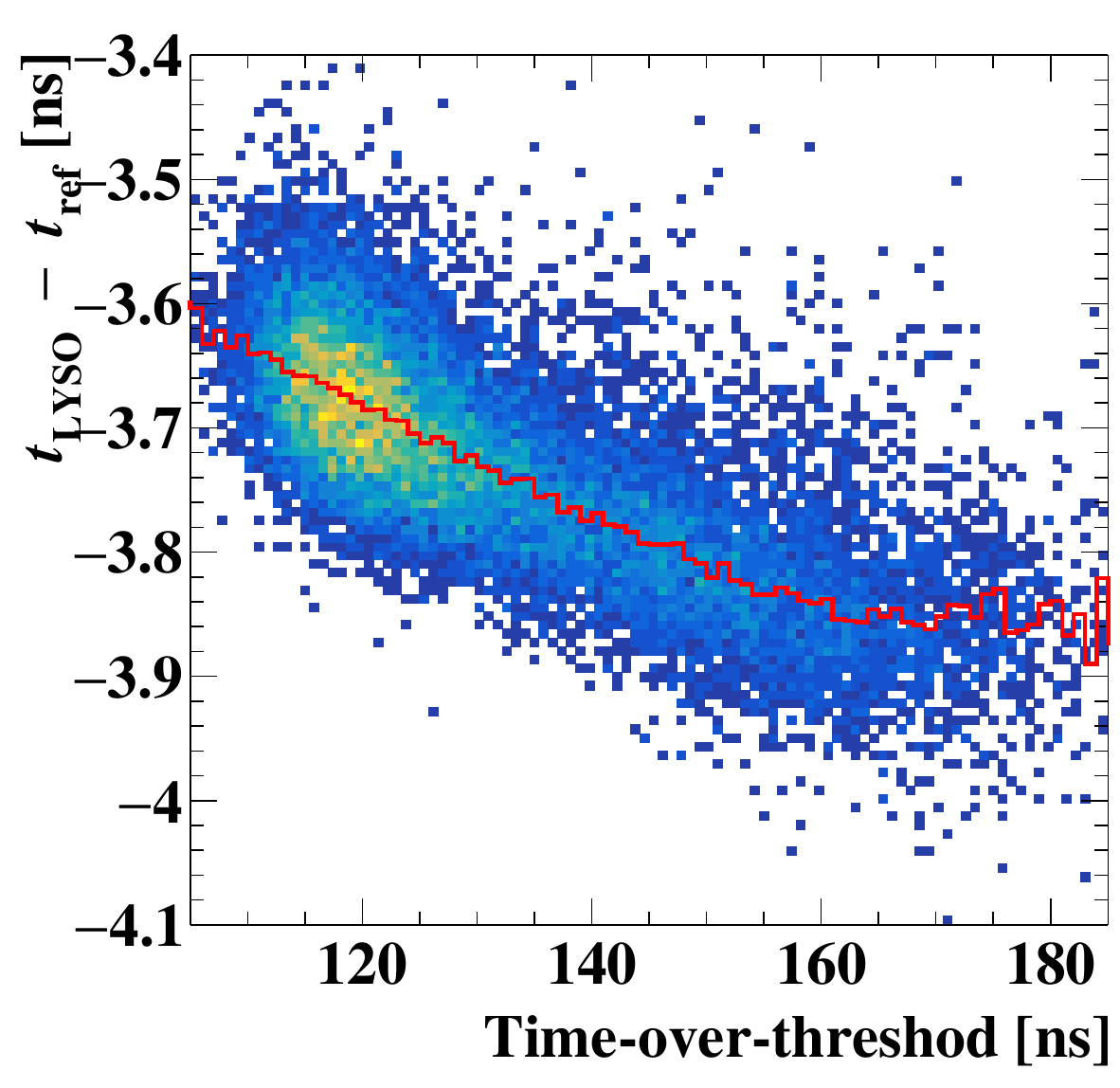}
      \subcaption{2023 data} % run453
      \label{fig:ToTTimeCorrelation}
   \end{minipage}
    \caption{
        Time-walk calibration based on the correlation between time offset and (a) low-gain charge or (b) high-gain time-over-threshold. 
        Correction functions are overlaid as the red lines.
    } 
   \label{fig:CorrelationPlots}
\end{figure}

\begin{figure}[tbp]
   \centering
   \begin{minipage}[t]{0.48\linewidth}
      \centering
      \includegraphics[width=1.\linewidth]{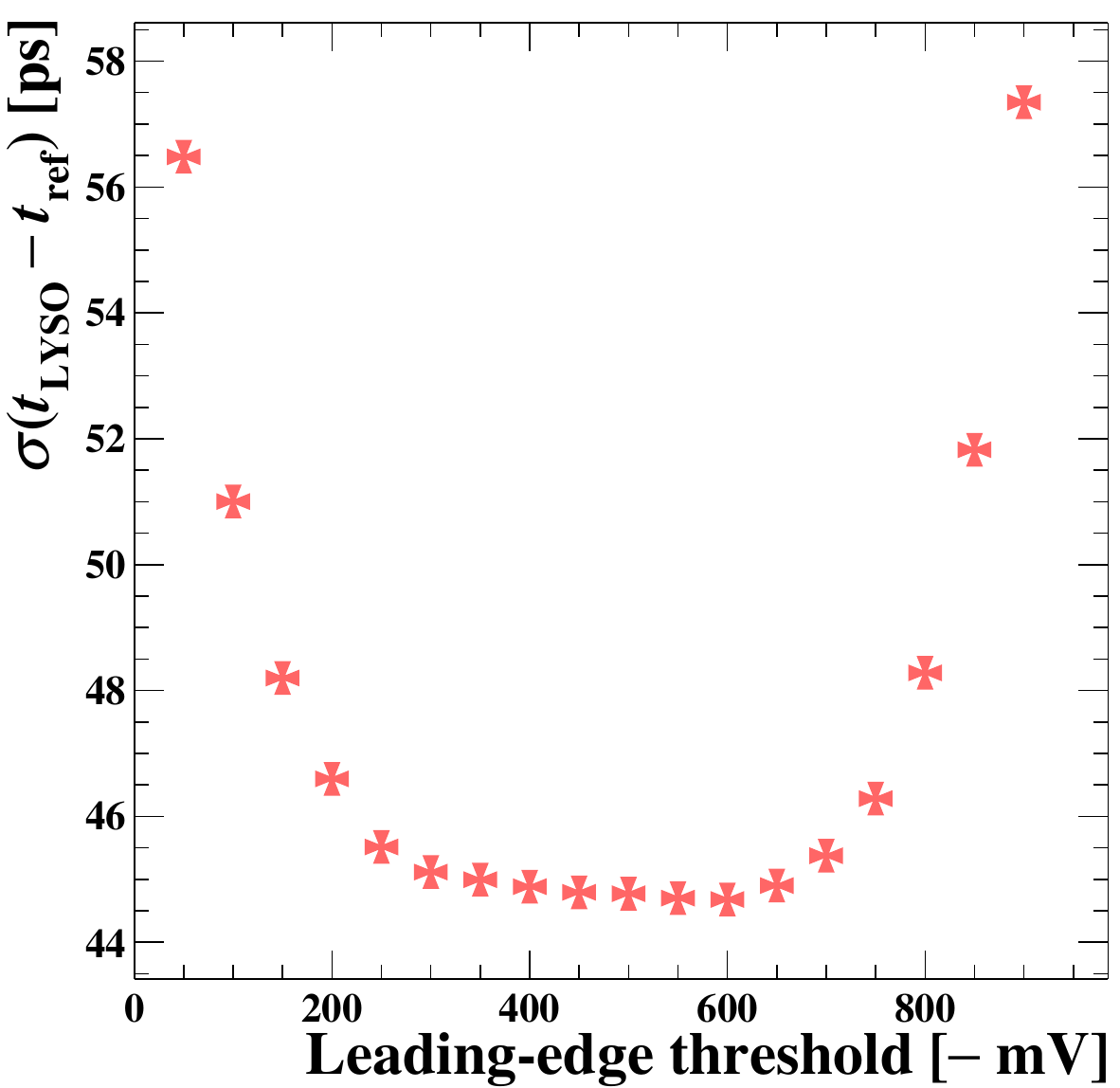}
      \subcaption{2024 data}
      \label{fig:thresholdOptimization2024}
   \end{minipage}
   \begin{minipage}[t]{0.48\linewidth}
      \centering
      \includegraphics[width=1.\linewidth]{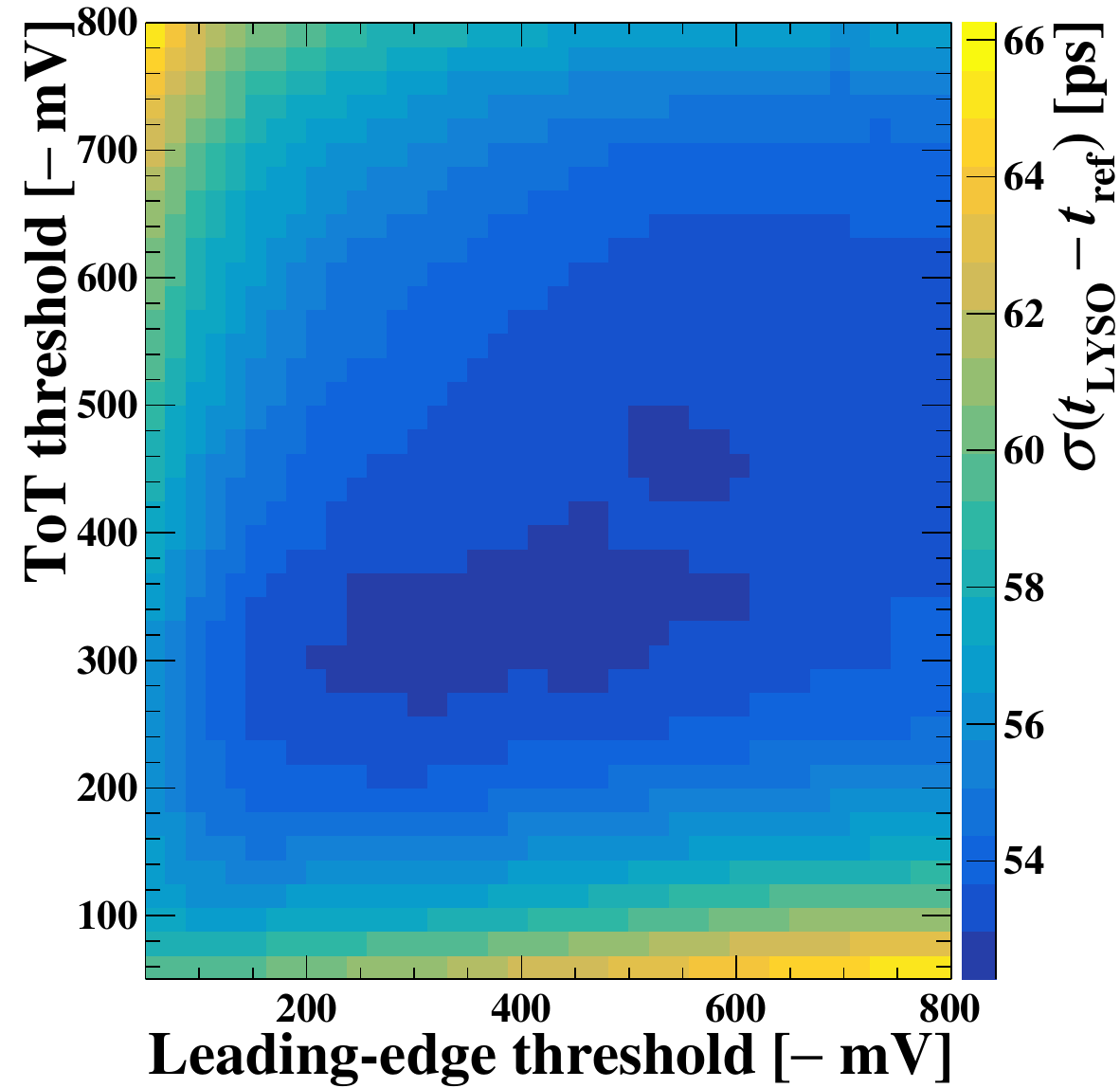}
      \subcaption{2023 data}
      \label{fig:thresholdOptimization2023}
   \end{minipage}
   \caption{
      (a) Time resolution as a function of the leading-edge threshold for the 2024 data.
      (b) Same measurement for the 2023 data, with the ToT dependence used for the walk correction.
      $t_\mathrm{LYSO}$ and $t_\mathrm{ref}$ denote the channel timing and the reference time, respectively.
      The optimal threshold (\SI{-600}{mV} for 2024) was used for the time resolution evaluation.
   }
   \label{fig:thresholdOptimization}
\end{figure}

\noindent Time calibration involved correcting the time-walk effect and optimizing the leading-edge threshold.
The time-walk effect was calibrated using the correlation between the leading-edge time and the pulse charge; specifically, we used the charge measured in the low-gain channel for the 2024 data (\figref{fig:ChargeTimeCorrelation}) and the ToT in the high-gain channel for the 2023 data (\figref{fig:ToTTimeCorrelation}).
As noted in \secref{sec:ReadOutElectronics}, the 2023 beam-test data were recorded only with high-gain settings.
The leading-edge threshold (and the ToT threshold for the 2023 data) was scanned to determine the optimal time resolution, as shown in \figref{fig:thresholdOptimization2024} and \figref{fig:thresholdOptimization2023}.

\subsection{Evaluation method of time resolution}\label{sec:TimeResolutionEvaluationMethod}
\noindent For configurations where each side has a single readout channel (either a single SiPM or three SiPMs in series connection; see \secref{sec:ActiveConverterPrototype} and \tabref{tab:SiPM}), the detection times for the left and right channels, 
$\tleft$ and $\tright$, were calculated directly from the leading edge of each waveform.
For configurations with three readout channels on each side, $\tleft$ and $\tright$ were obtained from the weighted average of the leading-edge times of the three channels. The weights were determined based on the resolution of each individual channel.
Typically, the central channel was assigned a weight approximately three times larger than those of the peripheral channels, as its sensitive area was fully contained within the LYSO end face.
\begin{figure}[tbp]
    \centering
    \includegraphics[width = 0.7\linewidth]{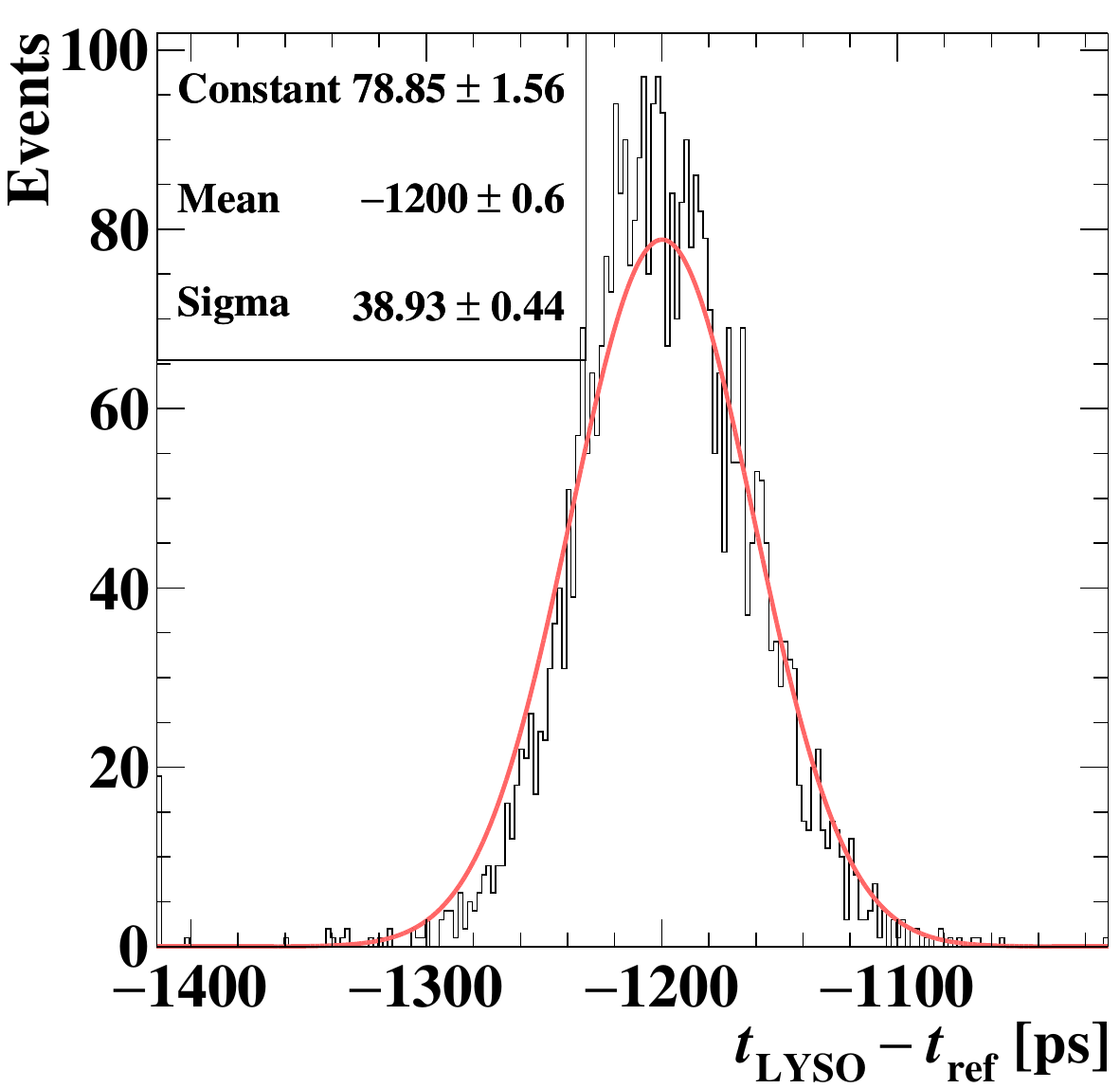}
    \caption{
       Time difference between the upstream \prototype~and the reference counters, measured using configuration~\rowref{row:20243mmSiPM} in \tabref{tab:Configurations}.
   }
    \label{fig:timeDistribution}
\end{figure}
\par
\indent The counter time resolution was obtained from the time differences among $t_\mathrm{US~LYSO}$, $t_\mathrm{DS~LYSO}$, and $t_\mathrm{ref}$ by solving the following system of equations:
\begin{equation}
   \begin{dcases}
   \sigma(t_\mathrm{US~LYSO} - t_\mathrm{ref})      & = \sqrt{\sigma_{t_\mathrm{US~LYSO}}^2 + \sigma_{t_\mathrm{ref}}^2}, \\
   \sigma(t_\mathrm{DS~LYSO} - t_\mathrm{ref})      & = \sqrt{\sigma_{t_\mathrm{DS~LYSO}}^2 + \sigma_{t_\mathrm{ref}}^2}, \\
   \sigma(t_\mathrm{US~LYSO} - t_\mathrm{DS~LYSO}) & = \sqrt{\sigma_{t_\mathrm{US~LYSO}}^2 + \sigma_{t_\mathrm{DS~LYSO}}^2}.
   \end{dcases}\label{eq:timeResolutionEvaluation}
\end{equation}
Here, $t_\mathrm{US/DS~LYSO}$ denotes the timing measured by the upstream or downstream prototype, defined as the average of the left and right channels: $(\tleft + \tright)/2$.
The term $t_\mathrm{ref}$ represents the reference timing, calculated as the average of the upstream and downstream reference counters.
\figref{fig:timeDistribution} shows an example distribution of $t_\mathrm{US~LYSO} - t_\mathrm{ref}$ together with its Gaussian fit.
\par
\indent In this study, we focused on the converter time resolution for single-MIP events, as discussed in \secref{sec:EventSelection}.
To ensure that the evaluation accurately reflects the performance for a typical single MIP, the time resolution was evaluated using events whose light yield fell within the $\pm$FWHM range around the MPV of the charge distribution (\figref{fig:eventSelection}).
This range corresponds to the core of the single-MIP energy-loss distribution while excluding the high-charge tail, which may be contaminated by multi-MIP contributions.
\begin{figure}[tbp]
   \centering
   \includegraphics[width=0.7\linewidth]{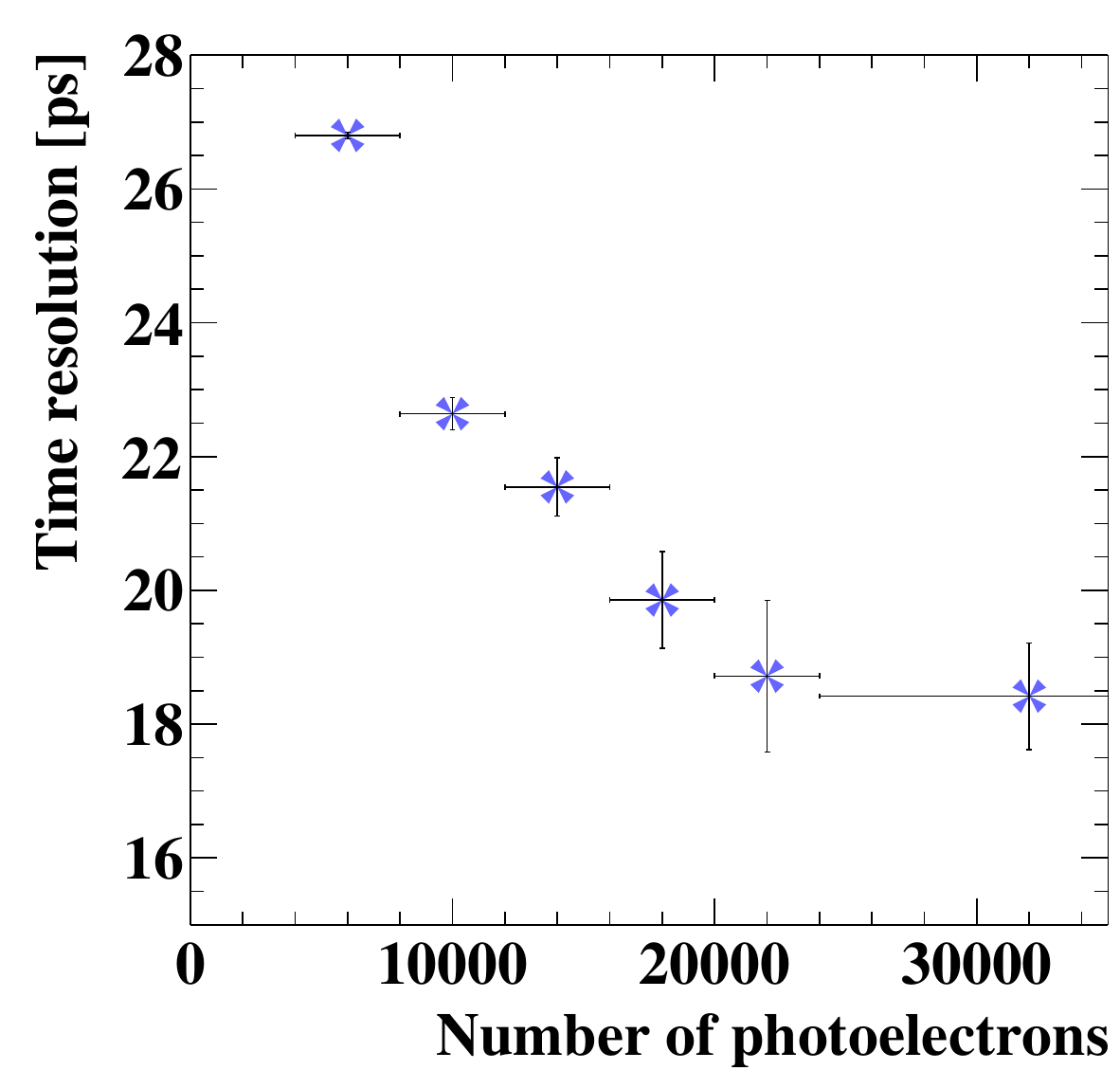}
   \caption{
      Time resolution of the {\prototype} at different ranges of the number of detected photoelectrons.
   }
   \label{fig:TimeResolution_LYdependence}
\end{figure}
Since the time resolution shows a clear dependence on the signal amplitude, as demonstrated in \figref{fig:TimeResolution_LYdependence}, restricting the sample to this well-defined light-yield window is essential for obtaining a representative and unbiased estimate of the performance.
\par
\indent As a cross-check to the $\sigma((\tleft + \tright)/2)$ evaluated above, the time resolution was also evaluated using the width $\sigma((\tleft - \tright)/2)$.
If $\tleft$ and $\tright$ are statistically independent and follow the same distribution, the Gaussian width of $(\tleft - \tright)/2$ should be equal to that of $(\tleft + \tright)/2$.
However, if there is a correlated effect between $\tleft$ and $\tright$, such as a residual time-walk effect, it would be canceled in the difference $(\tleft - \tright)$ but not in the sum $(\tleft + \tright)$.
In such cases, $\sigma((\tleft + \tright)/2)$ becomes larger than $\sigma((\tleft - \tright)/2)$.
In this study, these two estimates were found to be consistent within measurement uncertainty in most cases, indicating that the time-walk correction was effective.
Unless otherwise stated, the converter time resolution quoted in this paper refers to $\sigma((\tleft + \tright)/2)$.

\subsection{Hit position dependence}
\subsubsection{Time resolution}\label{sec:TimeResolutionPositionDependence}
\begin{figure}[tbp]
    \centering
    \includegraphics[width = 0.7\linewidth]{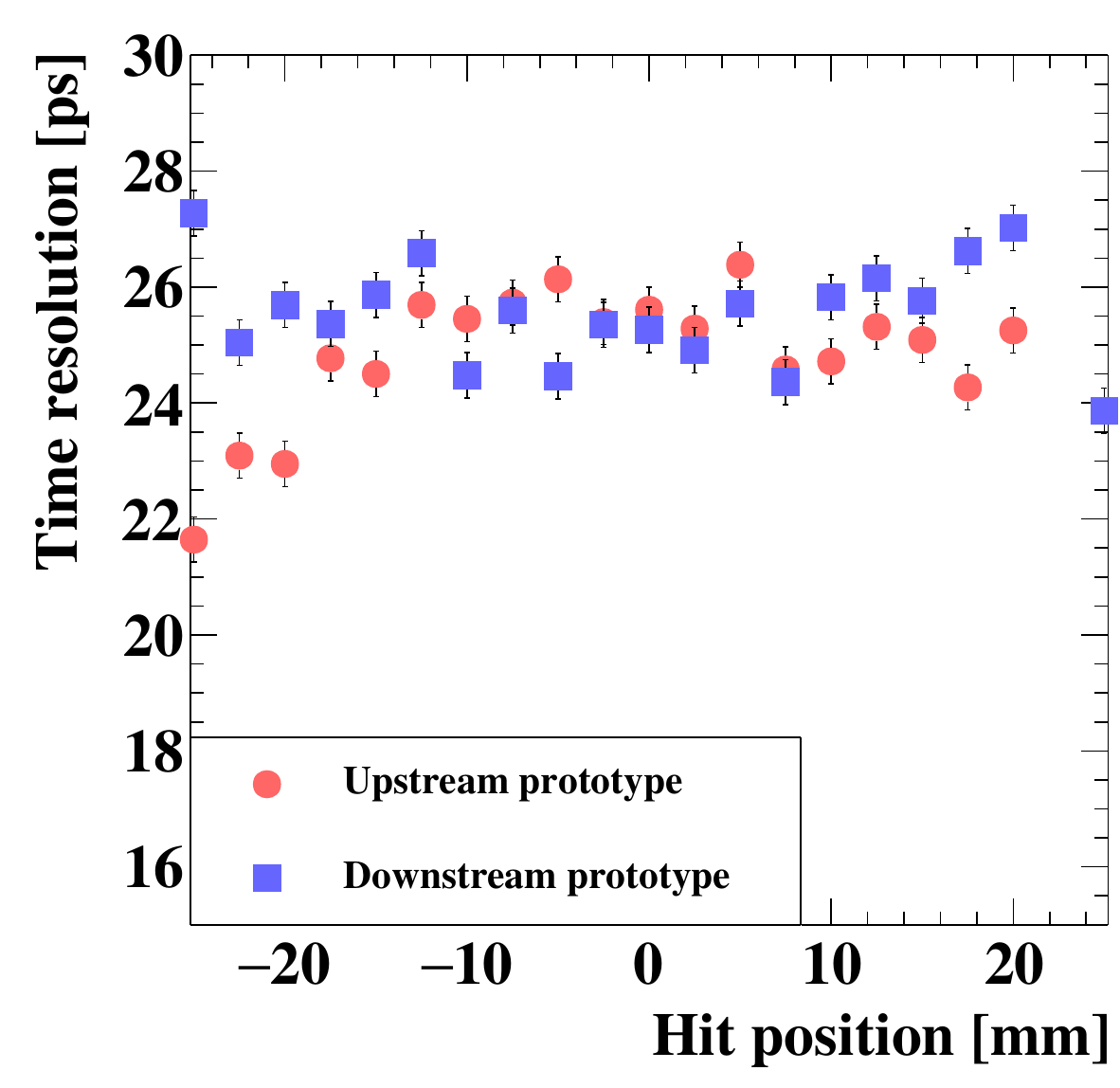}
    \caption{
       Time resolution of the upstream (downstream) LYSO crystal at different beam impact positions, 
       measured using configuration~\rowref{row:20243mmSiPM} in \tabref{tab:Configurations}, shown as red circles (blue squares).
   }
    \label{fig:positionScan_timeResolution}
\end{figure}

\noindent \figref{fig:positionScan_timeResolution} presents the measured time resolutions of the prototypes, positioned upstream and downstream along the beamline, as a function of the beam hit position.
For both counters, the resolution was found to be within the range of \SIrange{22}{27}{ps} over the entire length of the crystal.

\subsubsection{Time offset}\label{sec:TimeOffsetPositionDependence}
\begin{figure}[tbp]
    \centering
    \includegraphics[width = 0.7\linewidth]{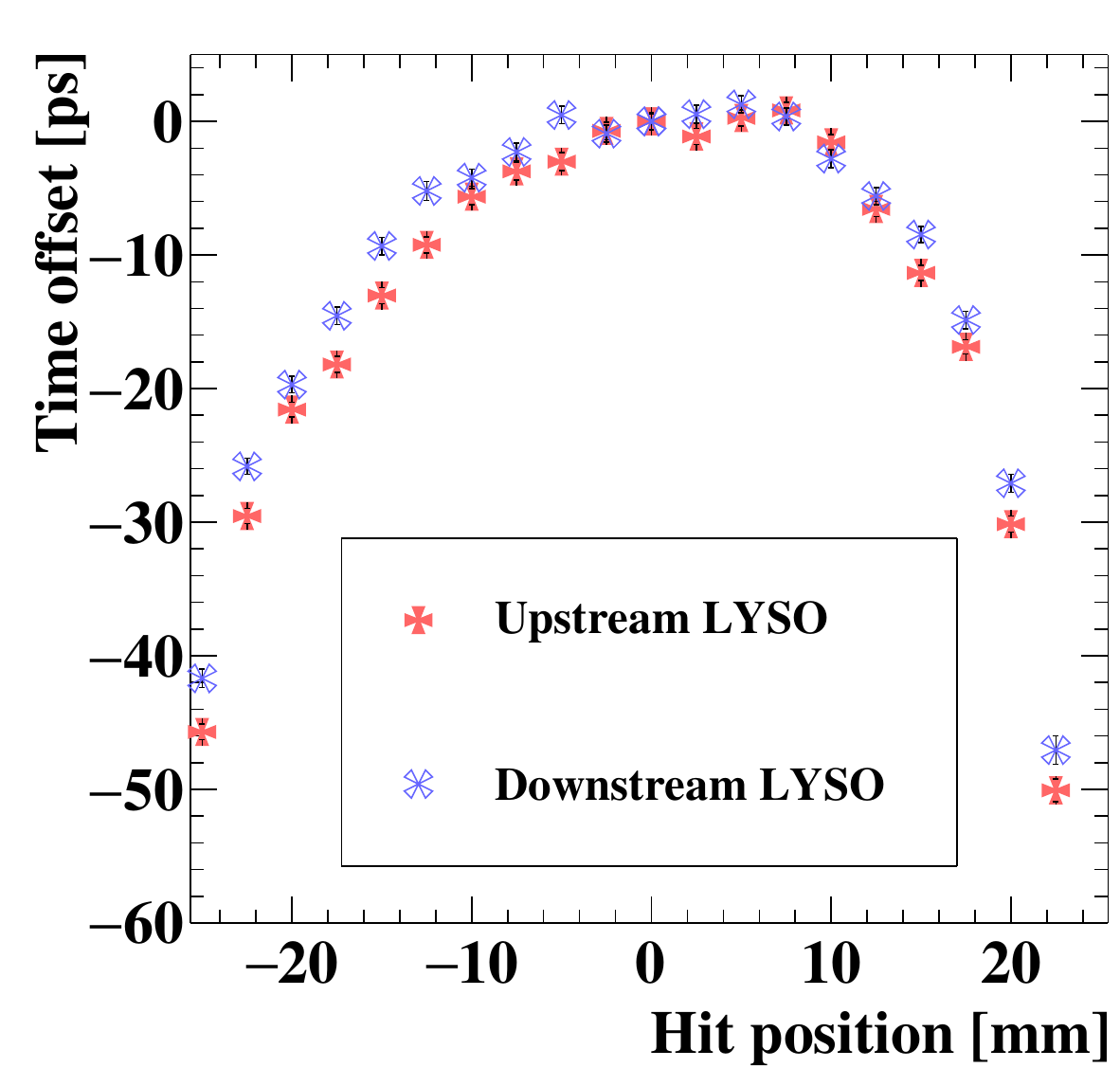}
    \caption{
          Offset of left-right averaged time at different beam impact positions, 
          measured with configuration~\rowref{row:20243mmSiPM} in \tabref{tab:Configurations}.
    }
    \label{fig:positionScan_timeOffsetAverage}
\end{figure}

\noindent A position-dependent variation of the time offset, up to \SI{50}{ps}, was observed within a single crystal, as shown in \figref{fig:positionScan_timeOffsetAverage}.
Once calibrated, however, this effect can be corrected using the entry positions of the reconstructed $e^+e^-$ tracks, leading to a negligible contribution to the overall \SI{25}{ps} resolution.
The counter-to-counter variation in this position dependence was found to be at most \SI{5}{ps}, which is also negligible compared to the total resolution.
\par
\begin{figure}[tbp]
    \centering
    \includegraphics[width = 0.7\linewidth]{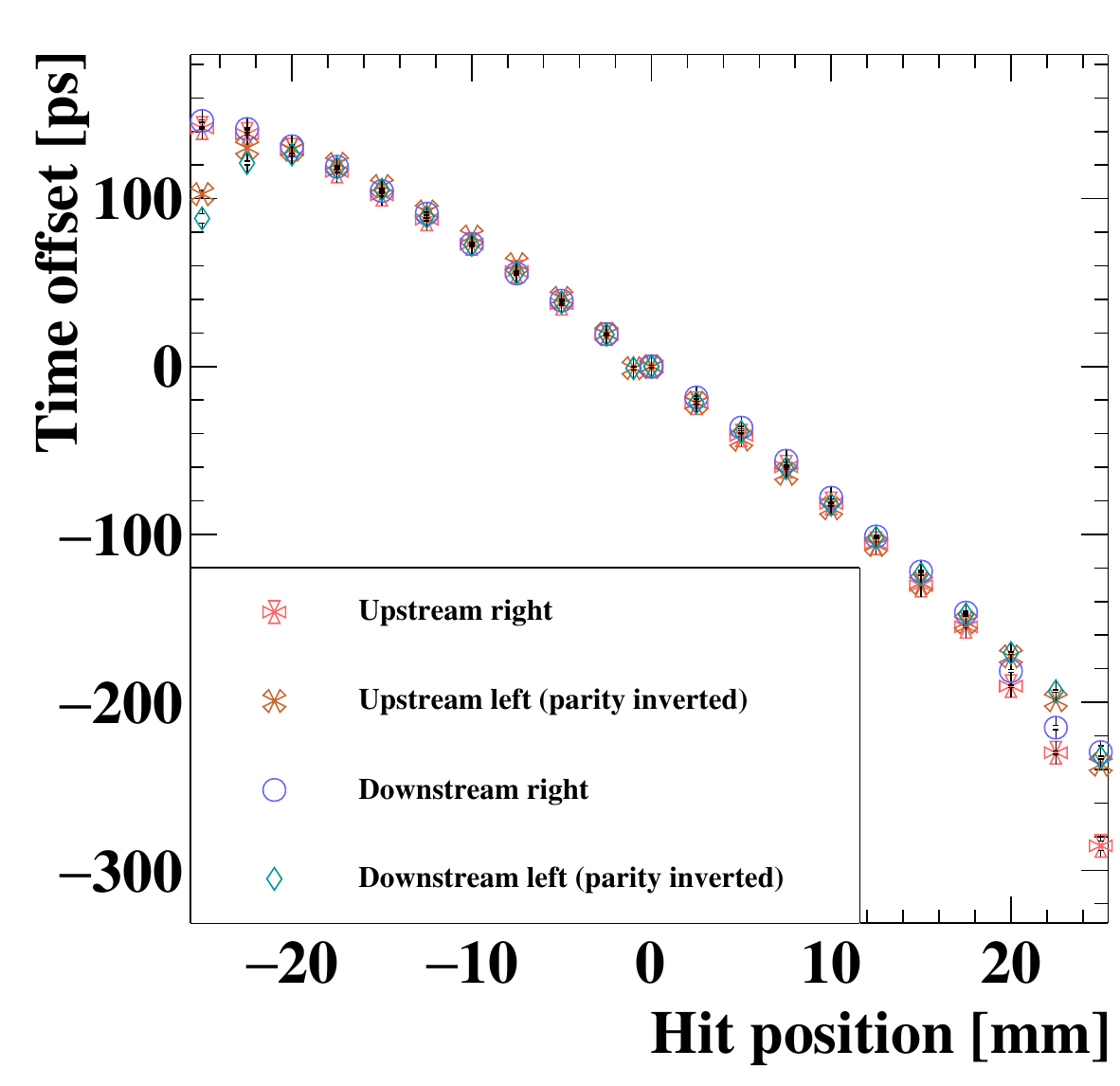}
    \caption{
         Time offset on each readout side at different beam impact positions, 
         measured with configuration~\rowref{row:20243mmSiPM} in \tabref{tab:Configurations}.
    }
    \label{fig:positionScan_timeOffsetOneside}
\end{figure}

\indent As single-sided readout is also being considered to reduce the number of channels in future experiments (see \secref{sec:fullScaleImplementation}), we investigated the position dependence of the time offset for individual readout channels.
\figref{fig:positionScan_timeOffsetOneside} compares four readout channels; note that the beam impact positions for the two right-side channels were parity-inverted with respect to the crystal center for comparison.
The observed channel-to-channel variation of up to \SI{10}{ps}, which is particularly pronounced near the crystal edges, could have a substantial impact on the resolution. This suggests that the single-sided scheme requires more careful calibration and quality control than the double-sided readout.

\subsection{Incident angle dependence}\label{sec:TimeResolutionAngleDependence}
\begin{figure}[tbp]
   \centering
   \includegraphics[width = 0.7\linewidth]{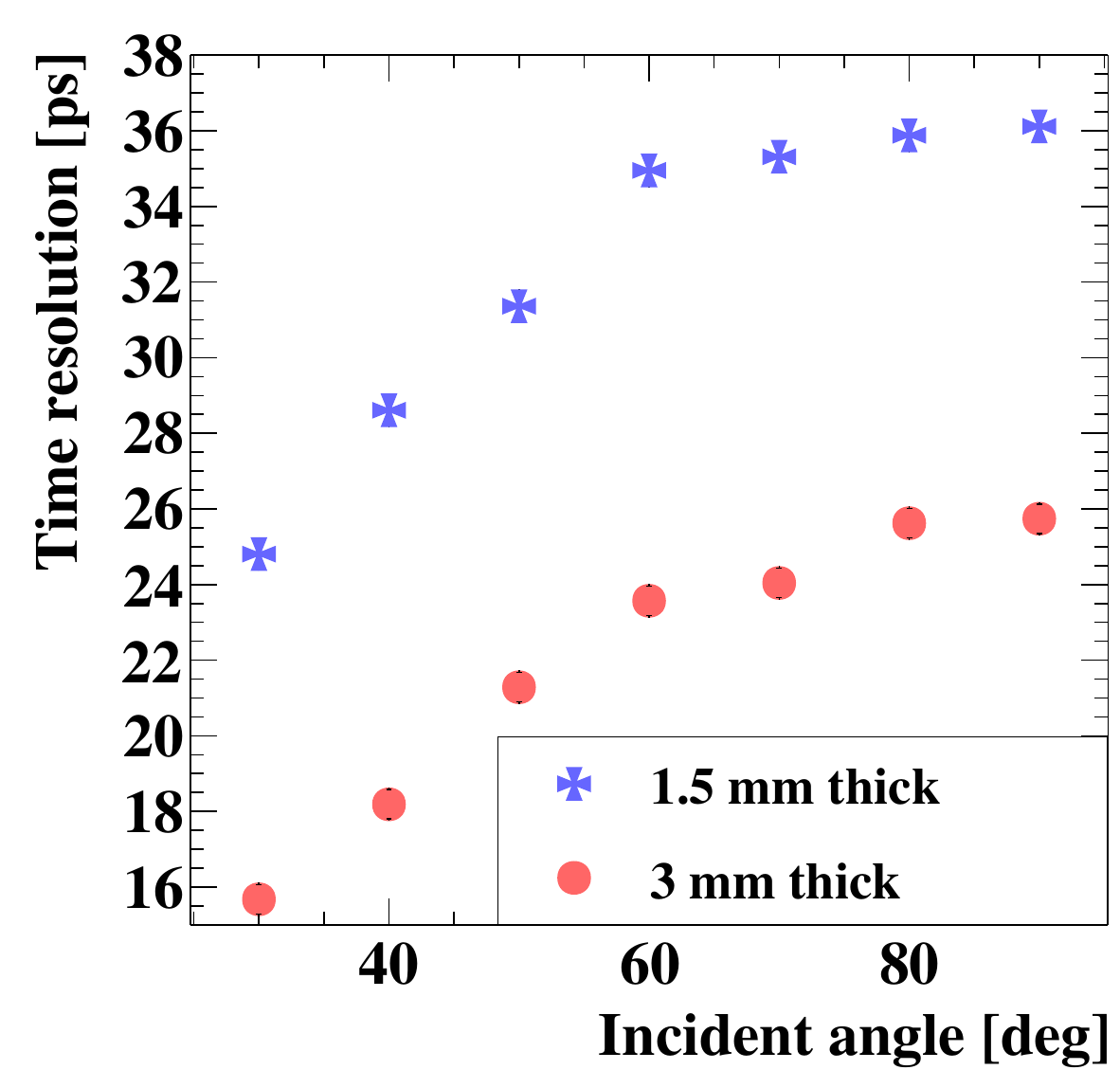}
    \caption{
       Time resolution at different beam incident angles for two different LYSO thicknesses, measured on the downstream side with 
       configuration~\rowref{row:20243mmSiPM}, \rowref{row:20246mmSiPMThinLYSO} in \tabref{tab:Configurations}.
    }
   \label{fig:angleScan_timeResolution}
\end{figure}

\noindent The time resolution of the prototype is shown in \figref{fig:angleScan_timeResolution} as a function of the beam incident angle relative to the converter surface.
The incident angle is defined in \figref{fig:angleScan_setup} to match the definition of $\theta_\gamma$ in \figref{fig:fullsketch}, where $\theta = \SI{90}{\degree}$ corresponds to perpendicular incidence.
The results presented here were obtained using the prototype positioned on the downstream side of the beamline.
As the incident angle decreases, the path length inside the crystal increases; consequently, the light yield increases, leading to an improvement in the time resolution.

\subsection{Performance with a thinner LYSO}
\noindent We also investigated the timing performance of prototypes with a reduced thickness, as thinner converters are considered for the outer regions of the detector (see \secref{sec:MaterialAndThickness}).
\figref{fig:angleScan_timeResolution} shows the performance of a \SI{1.5}{mm}-thick prototype evaluated using configuration~\rowref{row:20246mmSiPMThinLYSO} in \tabref{tab:Configurations}.
Although the time resolution was degraded compared to the \SI{3}{mm}-thick prototype due to the lower light yield, a resolution of \SI{27}{ps} was achieved at an electron incident angle of \SI{30}{\degree}. 
This corresponds to the typical photon incidence expected in the regions where \SI{1.5}{mm}-thick converters would be installed.
Even with perpendicular beam injection ($\theta = \SI{90}{\degree}$), a resolution of approximately \SI{35}{ps} was achieved across all impact positions, surpassing the target requirement of \SI{40}{ps}.

\subsection{SiPM types}\label{sec:TimeResolutionSiPMComparison}
\noindent The timing performance was compared among the three types of SiPMs listed in \tabref{tab:SiPM}, using configurations \rowref{row:20243mmSiPM}--\rowref{row:20244mmSiPM} in \tabref{tab:Configurations}.
For the S14160-6050HS, a single sensor with an active area of $6 \times \SI{6}{mm^2}$ was attached to each side of the crystal, providing full (\SI{100}{\percent}) coverage of the LYSO end face.
In the case of the S14160-3050HS, three $3 \times \SI{3}{mm^2}$ SiPMs were mounted on each side. 
This arrangement resulted in dead spaces between the devices, reducing the effective active-area coverage to approximately \SI{92}{\percent}; 
however, connecting the three SiPMs in series reduces the total capacitance, which is expected to enhance timing performance.
For the MICROFJ-40035, a single $4 \times \SI{4}{mm^2}$ SiPM was mounted on each side, limiting the end-face coverage to about \SI{80}{\percent}. 
This device, however, features a capacitively coupled ``fast output'' in addition to the standard output, providing a dedicated high-speed timing signal to mitigate the disadvantage of the smaller coverage.
\par
\begin{figure}[tbp]
   \centering
   \includegraphics[width = 0.7\linewidth]{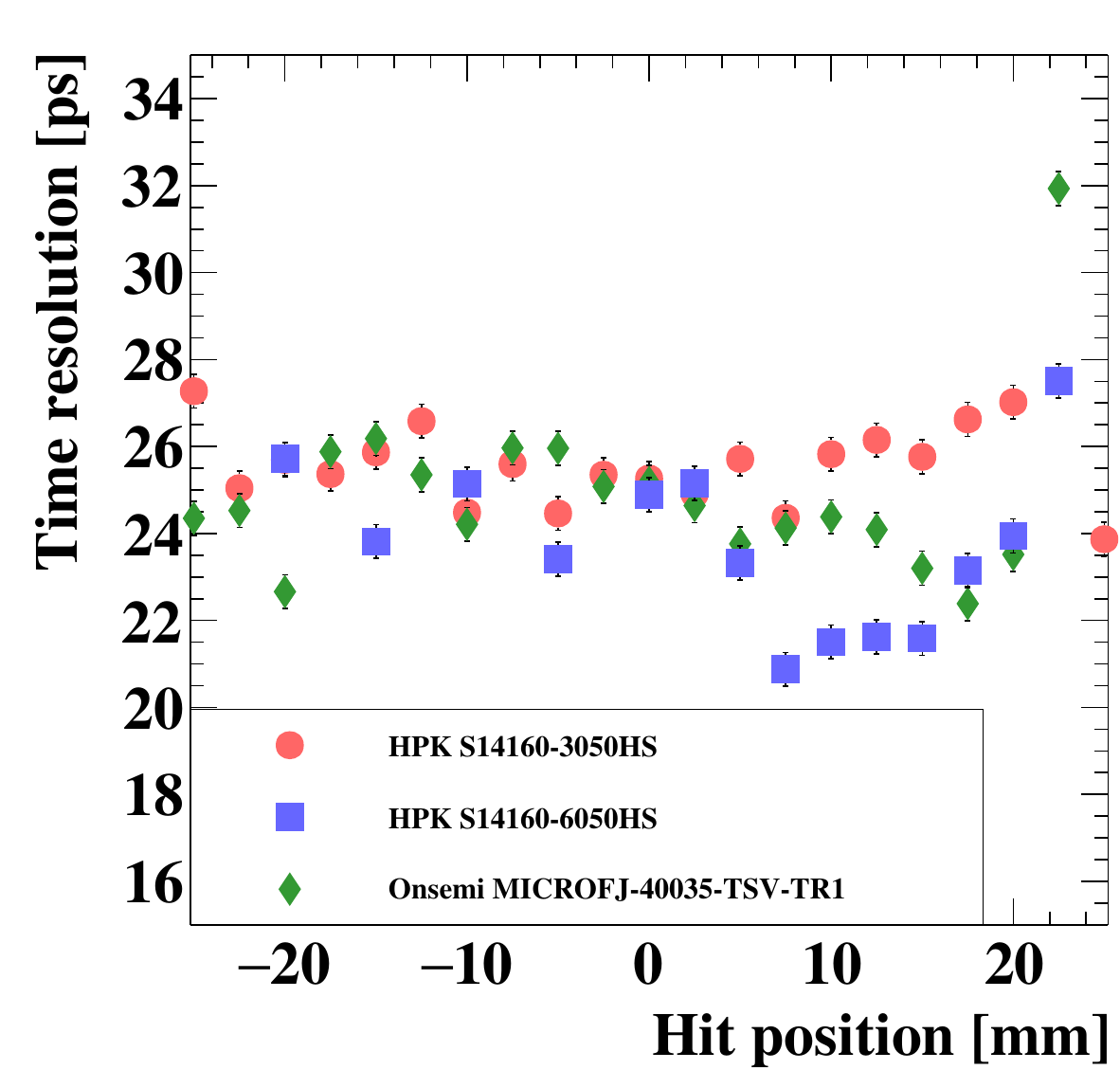}
   \caption{
      Time resolution with different types of SiPMs, listed in \tabref{tab:SiPM}, measured at different hit positions with configurations~\rowref{row:20243mmSiPM}--\rowref{row:20244mmSiPM} in \tabref{tab:Configurations}.
   }
   \label{fig:positionScan_SiPMComparison}
\end{figure}

\indent \figref{fig:positionScan_SiPMComparison} shows the time resolution of the prototypes obtained with each SiPM type as a function of the beam hit position.
Although differences in active-area coverage and readout schemes could in principle affect the timing performance, all three configurations achieved consistent time resolutions in the range of \SIrange{22}{27}{ps}, with no significant advantage observed for any particular SiPM model.

\subsection{SiPM readout method}
\noindent
The timing performance was compared between the two SiPM readout schemes (configurations \rowref{row:2023Series} and \rowref{row:2023Independent} in \tabref{tab:Configurations}), as summarized in \tabref{tab:ReadoutComparison}.
The large discrepancies between $\sigma((\tleft + \tright)/2)$ and $\sigma((\tleft - \tright)/2)$, which were observed only in the 2023 beam test, may be attributed to insufficient time-walk correction when using ToT instead of pulse charge (see \secref{sec:TimeResolutionEvaluationMethod}).
In the independent-readout and series-readout (single-readout) configurations, the values of $\sigma((\tleft - \tright)/2)$ were \SI{33}{ps} and \SI{27}{ps}, respectively.
The quantity $\sigma((\tleft - \tright)/2)$ represents the lower bound of $\sigma((\tleft + \tright)/2)$ even with perfect time-walk correction.
Consequently, the \SI{33}{ps} limit observed in the independent-readout configuration suggests that this scheme is less favorable for achieving high timing resolution.
Furthermore, the independent-readout scheme requires three times as many channels as the series-readout scheme.
Given these considerations of both timing performance and channel count, the series-readout scheme was adopted as the baseline configuration.

\begin{table}[tpb]
   \centering
   \begin{tabular}{ccc}\hline
                                       & \footnotesize Single readout & \footnotesize Independent readout \\\hline\hline
      $\sigma((\tleft + \tright)/2)$   & $39 \pm \SI{4}{ps}$      & $36 \pm \SI{4}{ps}$ \\\hline
      $\sigma((\tleft - \tright)/2)$   & $27 \pm \SI{2}{ps}$      & $33 \pm \SI{4}{ps}$ \\\hline
   \end{tabular}
   \caption{
      Time resolutions of the \prototype~for two readout schemes, measured using 
      configurations~\rowref{row:2023Series} and \rowref{row:2023Independent} in \tabref{tab:Configurations}.
      The electron beam was injected perpendicularly to the LYSO crystal center.
   }
   \label{tab:ReadoutComparison}
\end{table}

%%%%%%%%%%%%%%%%%%%%%%%%%%
%%                      %%
%% Light yield          %%
%%                      %%
%%%%%%%%%%%%%%%%%%%%%%%%%%
\section{Light yield}\label{sec:LightYield}
\subsection{Calibration}\label{sec:LightYieldCalibration}
% Photoelectron gain calibration
\noindent The light yield was evaluated by dividing the detected charge for MIP events by the single-photoelectron gain of the SiPMs.
The single-photoelectron gain was calibrated using the high-gain channels, which provided a sufficient signal-to-noise ratio to resolve individual photoelectron peaks.
However, since the charge for beam-induced signals was measured in the low-gain channels to avoid waveform saturation (as shown in \figref{fig:waveform}), it was necessary to calibrate the gain ratio between the high- and low-gain electronics branches.
Furthermore, a correction for the non-linear response of the SiPMs, arising from pixel saturation, was applied to the measured light yield.
% Single photoelectron gain
\begin{figure}[tbp]
    \centering
    \includegraphics[width = 0.7\linewidth]{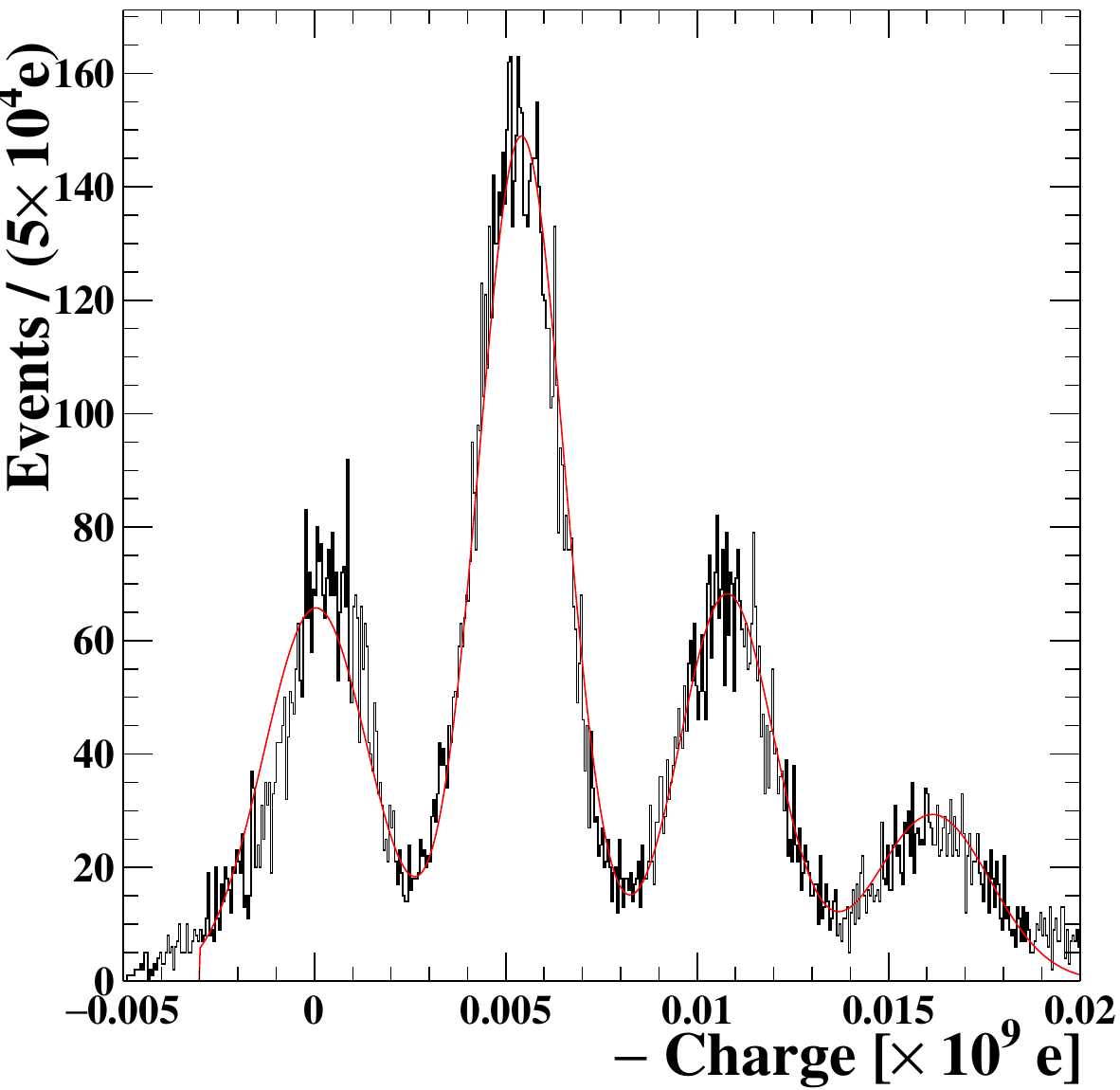}
    \caption{Charge distribution of SiPM when illuminated by weak LED light. 
    The peak corresponds to 0, 1, 2, and \SI{3}{photoelectrons} counted from left to right, respectively.}
    \label{fig:singlePhotoelectronChargeDistribution}
\end{figure}
\par
\indent The single-photoelectron gain was calibrated in the high-gain setting by resolving individual peaks in the charge distribution of the SiPMs illuminated with low-intensity LED light (\figref{fig:singlePhotoelectronChargeDistribution}).
These peaks were fitted using a sum of multiple Gaussian functions, with the gain extracted from the separation between adjacent peaks.
Since the ambient temperature during the beam test differed from that at the time of gain calibration, the gain was corrected using the temperature coefficient provided by the manufacturer.
\par
% WDB gain calibration
\indent Calibration of the amplifier gains integrated into the WaveDREAM board is necessary to relate the charge detected in the low-gain channels to the single-photoelectron gain calibrated in the high-gain channels.
To account for the frequency response of the electronics, this calibration was performed using an exponential pulse from a function generator with a time constant close to that of the SiPM signals.
As a result, the high-to-low gain ratio was determined to be $47.7$ with an uncertainty of \SI{2.4}{\percent}.
\par
\begin{figure}[tbp]
   \centering
   \includegraphics[width=0.7\linewidth]{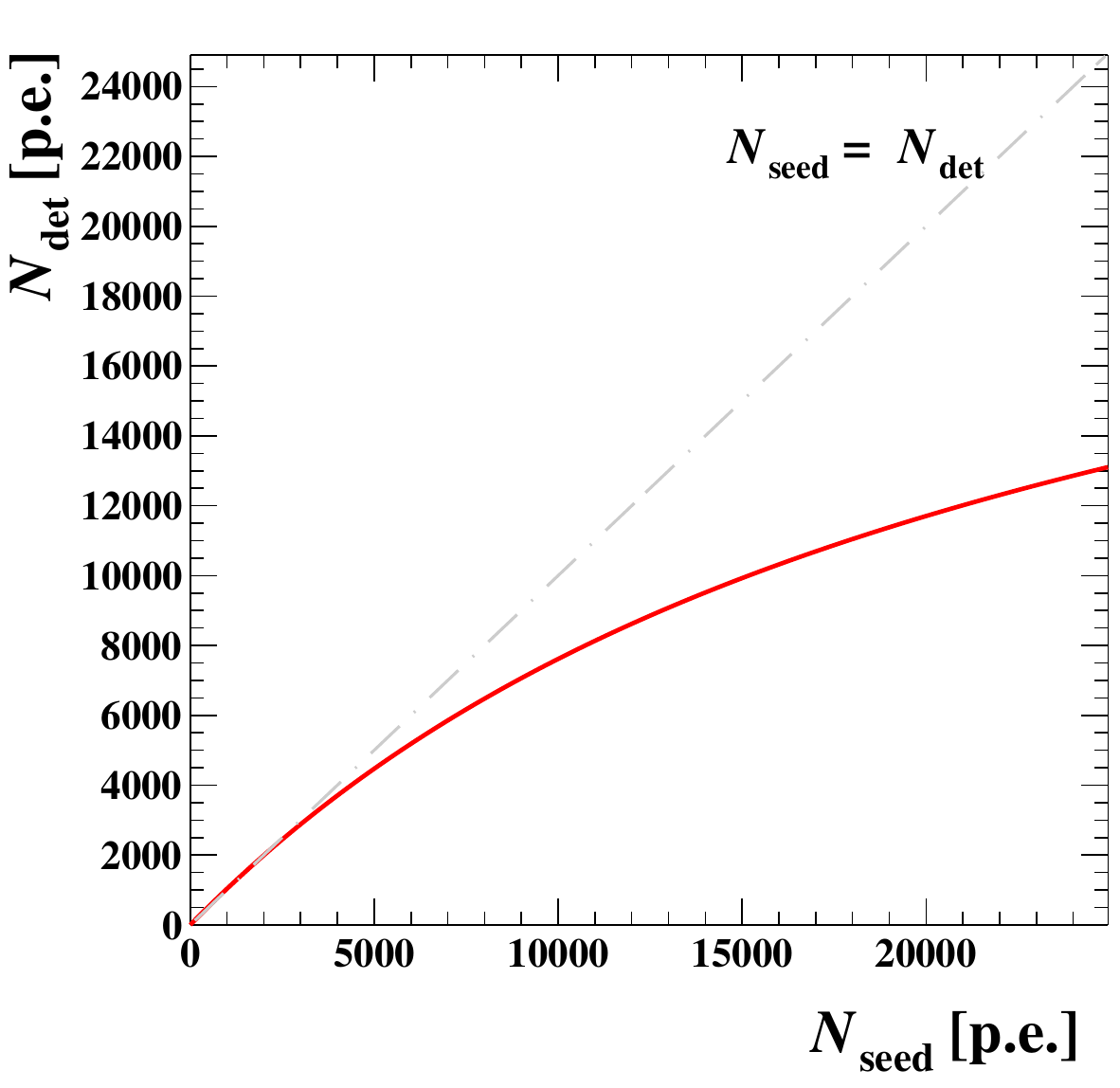}
   \caption{Saturation curve obtained from the model described in \cite{SiPMsaturation} for HPK S14160-3050HS.
   $N_\mathrm{seed}$ and $N_\mathrm{det}$ denote the number of photoelectrons in the absence of saturation and that expected to be detected by the SiPM, respectively.}
   \label{fig:SaturationCurve}
\end{figure}

% SiPM non-linearity correction
\indent A correction for the non-linear response of the SiPMs was also applied, based on the saturation model described in \cite{SiPMsaturation}.
This model accounts for the relationship between the expected number of photoelectrons in the absence of saturation, $N_\mathrm{seed}$, and the observed number of detected photoelectrons, $N_\mathrm{det}$, as illustrated in \figref{fig:SaturationCurve}.
The SiPM parameters required for the model, such as the optical cross-talk probability and pixel recovery time, were taken from the specifications provided by the manufacturer.

\subsection{Evaluation method for light yield}
\begin{figure}[tbp]
    \centering
    \includegraphics[width = 0.7\linewidth]{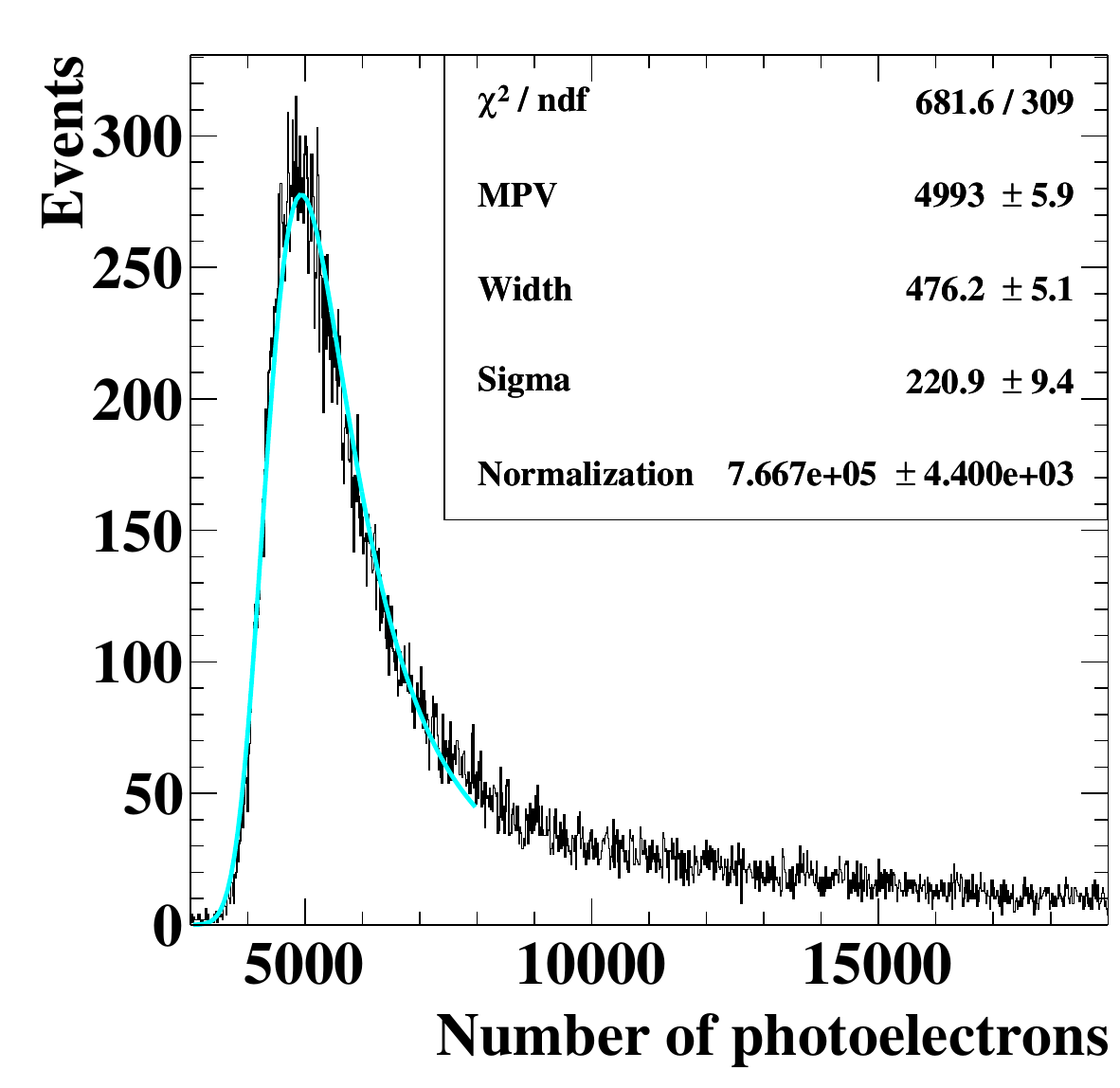}
    \caption{
    Distribution of the number of photoelectrons detected at the right end of the crystal bar, with configuration~\rowref{row:20243mmSiPM} in \tabref{tab:Configurations}.}
    \label{fig:lightYieldDistribution}
\end{figure}

\noindent The distribution of the number of photoelectrons detected on each side of the crystal was fitted with a Landau function convoluted with a Gaussian function; an example is shown in \figref{fig:lightYieldDistribution}.
In this paper, the light yield of the prototype is defined as the sum of the light yields measured on both sides of the crystal.

\subsection{Hit position dependence}\label{sec:LightYieldPositionDependence}
\begin{figure}[tbp]
    \centering
    \includegraphics[width = 0.7\linewidth]{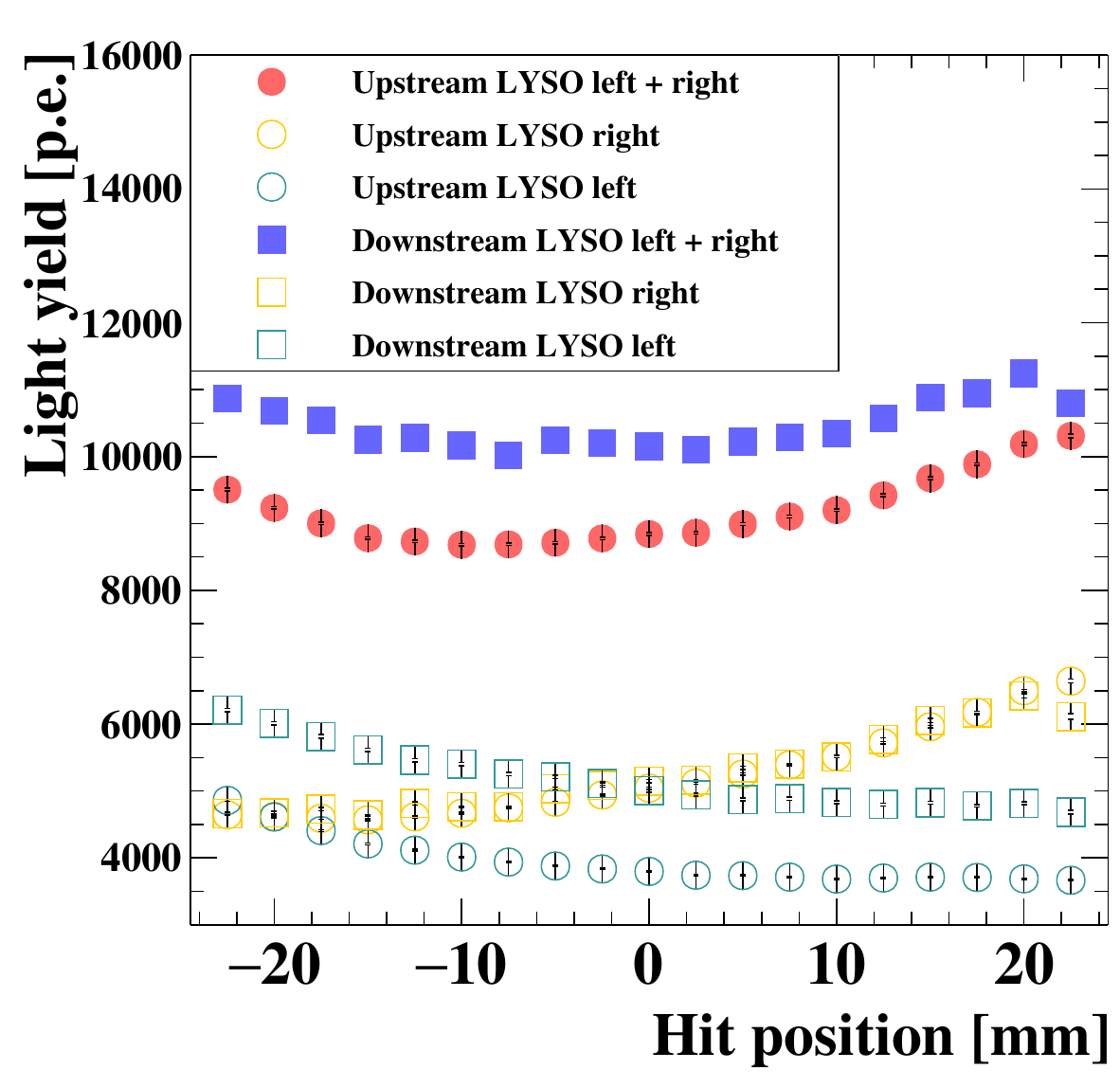}
    \caption{
       Light yield measured at different beam impact positions using configuration~\rowref{row:20243mmSiPM} in \tabref{tab:Configurations}. 
    }
    \label{fig:positionScan_lightYield}
\end{figure}

\noindent The light yields of the upstream and downstream prototypes at various hit positions are presented in \figref{fig:positionScan_lightYield}.
While the total light yield, on the order of \num{e4} photoelectrons, exceeds the requirement, a decrease in light yield is observed near the central region of the crystal bar, with an overall variation of approximately \SI{15}{\percent}.
To suppress its contribution to the energy resolution below the target level, this position dependence must be calibrated and corrected using the reconstructed photon impact position.
\par
\indent In addition to the position-dependent variation discussed above, a notable difference was observed between the two prototypes. For the downstream counter, the light yields measured on both sides were consistent with each other at the center; however, the upstream counter exhibited a pronounced asymmetry.
This asymmetry may originate from factors such as imperfect optical coupling, non-uniform wrapping, or other conditions affecting the light collection efficiency of the upstream left channel. This suggests that the assembly quality significantly influences the light yield.

\subsection{Incident angle dependence}\label{sec:LightYieldAngleDependence}
\begin{figure}[tbp]
   \centering
   \includegraphics[width = 0.7\linewidth]{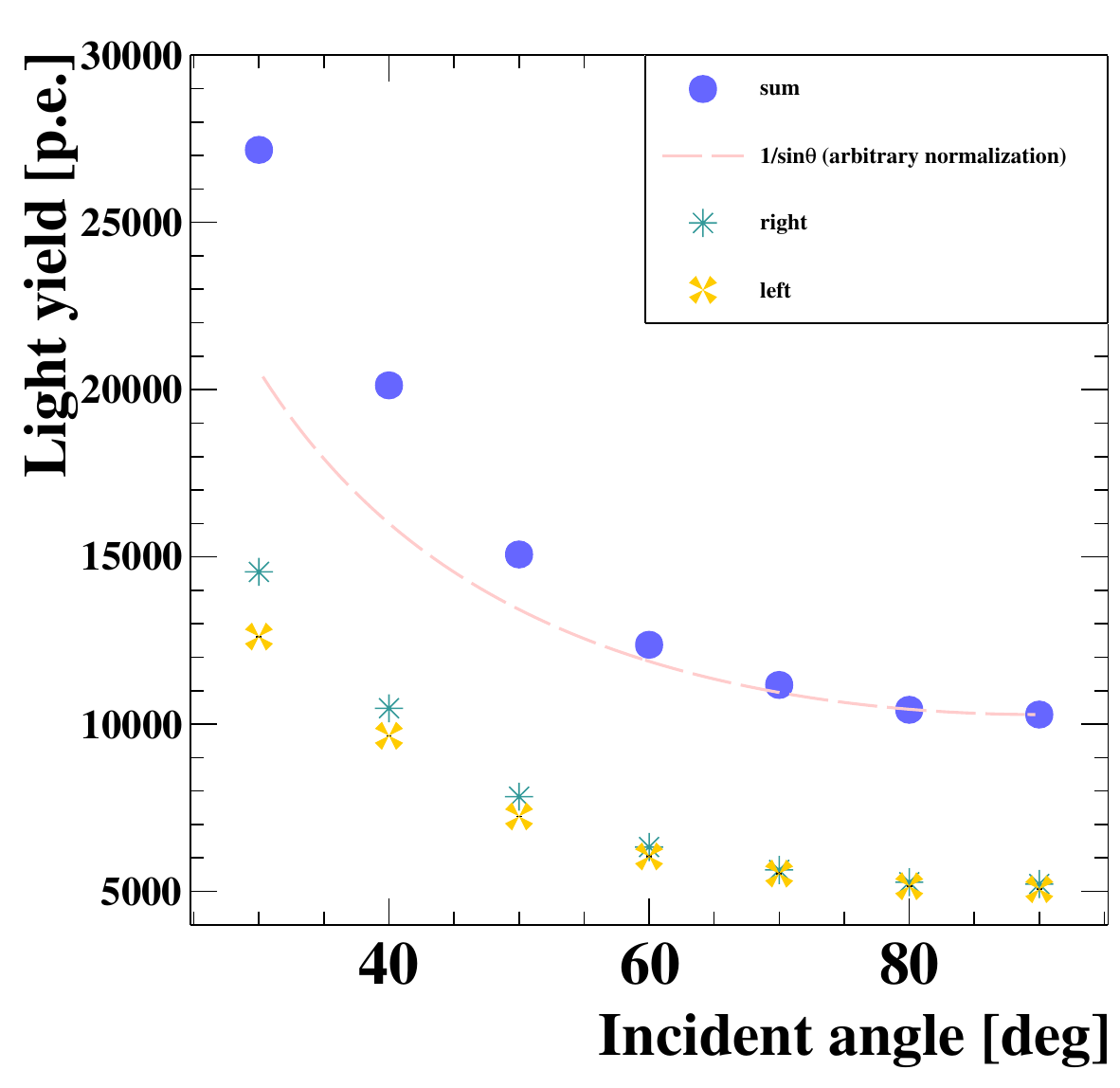}
    \caption{
       Light yield at different beam incident angles with the configuration~\rowref{row:20243mmSiPM} in \tabref{tab:Configurations}, measured with the downstream counter.
    }
   \label{fig:angleScan_lightYield}
\end{figure}

\noindent The light yield as a function of the beam incident angle for the downstream prototype is shown in \figref{fig:angleScan_lightYield}.
The observed trend is qualitatively consistent with the $1/\sin\theta$ dependence expected from the path length of the electron beam within the LYSO crystal. Some deviations from this ideal behavior are observed at smaller incident angles, likely due to effects such as edge effects or light collection non-uniformity.

%%%%%%%%%%%%%%%%%%%%%%%%%%
%%                      %%
%% Discussion           %%
%%                      %%
%%%%%%%%%%%%%%%%%%%%%%%%%%
\section{Discussion}\label{sec:Discussion}
\subsection{Performance of LYSO as an active converter}\label{sec:OverallLYSOPerformance}
\noindent The timing performance of the LYSO active converter, measured under various conditions, consistently demonstrated high precision.
At a thickness of \SI{3}{mm}, the typical time resolution was approximately \SI{25}{ps}.
Compared to the design requirement of \SI{40}{ps} (see \secref{sec:DesignConcept}), these results exhibit a significant performance margin.
Across all tested parameters
---
including beam impact position (\secref{sec:TimeResolutionPositionDependence}), 
incident angle and thickness (\secref{sec:TimeResolutionAngleDependence}), and SiPM/readout configuration (\secref{sec:TimeResolutionSiPMComparison})
---
the measured resolutions remained well within the \SI{40}{ps} limit.
These findings confirm that a total photon time resolution of \SI{30}{ps} is achievable for the conversion spectrometer by combining the electron and positron timing.
\par
\indent Regarding the scintillation light output, the measured light yield for a typical MIP energy deposit (\SI{3.5}{MeV}) reached the order of \num{e4} photoelectrons.
This value is more than an order of magnitude higher than the minimum requirement of \SI{700}{photoelectrons} derived in \secref{sec:ConverterLYrequirement} 
(which corresponds to \SI{2500}{photoelectrons} for the maximum expected energy deposit of \SI{10}{MeV} from the $e^+e^-$ pair).
As demonstrated in the position scan (\secref{sec:LightYieldPositionDependence}), the light yield remained consistently above the threshold, 
ensuring that statistical fluctuations are suppressed below $\sim$\SI{50}{keV}.
Consequently, the target energy resolution of \SI{200}{keV} will not be limited by photon statistics, but rather by the $e^+e^-$ tracking precision or calibration quality.

\subsection{Prospects for the full-scale implementation of a conversion spectrometer}\label{sec:fullScaleImplementation}
\noindent Regarding the calibration of a full-scale detector, our studies identified specific variations that define the necessary calibration precision.
Specifically, we observed a position-dependent variation of approximately \SI{50}{ps} in the time offset and \SI{15}{\percent} in the light yield within a single crystal.
To maintain the target performance across the entire converter, these variations must be corrected using the hit positions reconstructed by $e^+e^-$ tracking.
While crystal-to-crystal timing differences were found to be below \SI{10}{ps}
---indicating they are not a primary concern after initial calibration---
we observed light-yield variations of up to a factor of 1.3 between different crystals.
These results highlight that achieving high uniformity in a full-scale system will depend on establishing well-controlled assembly procedures and robust, 
precise calibration methods.
\par
\indent Another major concern for a full-scale detector is the large number of readout channels. 
For example, in the four-layer configuration illustrated in \figref{fig:fullsketch}, the total number of converter segments is estimated to be on the order of \num{e5}. 
If single-sided readout were to be employed, the channel count could be halved. 
Although the single-sided time resolution is approximately a factor of $\sqrt{2}$ larger than that of the double-sided configuration, 
the measured value of about \SI{35}{ps} still satisfies the design requirement.
However, position-dependent variations and crystal-to-crystal differences become more prominent with single-sided readout 
(see \secref{sec:TimeOffsetPositionDependence}), and further studies are required to address these effects.
An alternative concept has also been proposed in which the signals from the left- and right-side channels are electrically combined 
and read out as a single channel. 
This approach aims to reduce the channel count while preserving the superior timing and light-yield performance of double-sided collection. 
Further investigation is necessary to assess the hardware feasibility and long-term stability of this scheme.
\par
\indent Accommodating a large number of converter segments also presents significant mechanical challenges. 
The estimated total mass of several hundred kilograms can be managed by modularizing the converter into sectors of approximately \SIrange{1}{2}{kg} each. 
Simulation studies confirm that the thin support structures required for these modules do not significantly degrade the \SI{52.8}{MeV} signal peak.
Regarding thermal constraints, the heat dissipation from the SiPMs is expected to be sufficiently low 
and is not anticipated to pose a significant issue. 
Furthermore, since the front-end readout electronics are located outside the detector volume, 
local heat accumulation is effectively mitigated. 
This indicates that the thermal load within the detector remains minimal, allowing for the implementation 
of conventional cooling systems.
\par
\indent Finally, while this study focused on the performance of the LYSO converter, the overall spectrometer performance will also depend on the pair-tracker.
Achieving the target \SI{85}{\percent} tracking efficiency and \SI{200}{keV} momentum resolution remains a critical requirement, 
which must be addressed through dedicated studies of the tracking system in the future.

\subsection{Comparison with previous studies using LYSO crystals}
\noindent In \citeref{CMSMTD:2021imi,CMS:Addesa:2024mpu,CMS2026}, LYSO bars with dimensions of $3.75\times 3.2\times \SI{54.7}{mm^3}$ were investigated for the CMS MIP Timing Detector, demonstrating a time resolution of \SI{25}{ps} for non-irradiated modules.
Motivated by the requirements of future $\mu \to e \gamma$ experiments, we identified an optimized LYSO geometry of \lysodimension{50}{5}{3} as our baseline through dedicated simulations and subsequently characterized its performance in detail.
In addition to achieving a comparable time resolution, this work demonstrated a high light yield of the order of $10^4$ photoelectrons—a crucial property for the energy reconstruction required in $\mu \to e \gamma$ searches.
Furthermore, we evaluated the performance across a broader parameter space, including thinner \SI{1.5}{mm} crystals, various incident angles, and different SiPM types and readout schemes.
These comprehensive studies have successfully established the feasibility of the LYSO-based active converter design for high-precision spectroscopy.

%%%%%%%%%%%%%%%%%%%%%%%%%%
%%                      %%
%% Conclusion           %%
%%                      %%
%%%%%%%%%%%%%%%%%%%%%%%%%%
\section{Conclusion}\label{sec:Conclusion}
\noindent This paper has presented a comprehensive performance evaluation of LYSO as an active converter material for a conversion spectrometer, a candidate detector technology for future \meg searches.
Through simulation studies focusing on conversion efficiency, pile-up capability, and background suppression, an optimized design with a baseline segmentation of \lysodimension{50}{5}{3} was identified.
The timing resolution and light yield of the LYSO prototype were characterized under various conditions
---including beam impact position, incident angle, and SiPM configuration---
using the KEK Test Beam Line with a \SI{3}{GeV} electron beam.
\par
\indent The experimental results demonstrated a typical single-MIP time resolution of \SI{25}{ps} and a light yield of the order of $10^4$ photoelectrons, 
both of which significantly exceed the design requirements of \SI{40}{ps} and \SI{700}{photoelectrons}, respectively.
For \SI{52.8}{MeV} photon detection, these results predict an achievable time resolution better than \SI{30}{ps}. 
While the overall spectrometer performance will include additional contributions from the pair-tracker and calibration uncertainties, the contribution from LYSO photoelectron statistics alone was estimated to be negligible at below \SI{50}{keV}.
In conclusion, this study has successfully established the LYSO-based active converter as a highly promising technology for the conversion spectrometer concept in next-generation \meg experiments.

\section*{Acknowledgement}
This work was supported by JSPS KAKENHI Grant Numbers JP26000004, JP20H00154, JP21H04991, JP21H00065, and JP22K21350.
The authors would like to express their gratitude to the staff of the KEK PF-AR test beamline for their excellent support and cooperation during the beam tests.

\bibliography{mybibfile}

\end{document}